\documentclass[a4paper,11pt]{article}
\pdfoutput=1 

\usepackage{jcappub} 

\usepackage[T1]{fontenc} 

\newcommand{\beq}{\begin{eqnarray}}
\newcommand{\eeq}{\end{eqnarray}}

\newcommand{\br}{\mathcal{B}}

\newcommand{\vescE}{{v_{\rm esc}^{\oplus}}}

\title{Direct and indirect detection of dissipative dark matter}

\author[a,b]{JiJi Fan,}
\author[a]{Andrey Katz,}
\author[a]{and Jessie Shelton}
\affiliation[a]{Department of Physics, Harvard University, Cambridge, MA 02138 }
\affiliation[b]{Physics Department, Syracuse University, Syracuse, NY, 13244}

\abstract{We study the constraints from direct detection and solar capture on dark matter  scenarios with a subdominant dissipative component.  This dissipative dark matter component in general has both a symmetric and asymmetric relic abundance.
Dissipative dynamics allow this subdominant dark matter component to cool, resulting in its partial or total collapse  into a smaller volume inside the halo (e.g., a dark disk) as well as a reduced thermal velocity dispersion compared to that of normal cold dark matter. We first show that these features considerably relax the limits from direct detection experiments on the couplings between standard model (SM) particles and dissipative dark matter. On the other hand, indirect detection of the 
annihilation of the symmetric dissipative dark matter component inside the Sun sets stringent and robust constraints on the properties of the dissipative dark matter. In particular, IceCube observations force dissipative dark matter particles with mass above 50 GeV to either have a small coupling to the SM or a low local density in the solar system, or to have a nearly asymmetric relic abundance. Possible helioseismology signals associated with purely asymmetric dissipative dark matter are discussed, with no present constraints.
  }

\begin{document}
\maketitle
\flushbottom

\section{Introduction}
\label{sec:intro}

The existence of dark matter (DM) is well established by gravitational observations and serves as the most tantalizing evidence for physics beyond the
Standard Model (SM). Yet very little is known about the composition
and interactions of dark matter beyond the universal gravitational
interaction. A tremendous amount of theoretical effort has been
invested in scenarios with single-component, cold, collisionless dark
matter, a picture mainly inspired by large-scale structure
observations and theoretical simplicity. However, we know very little about the dark world and, as the visible world consists of
a complicated multi-scale combination of thermal and non-thermal relics, it is important to
explore less minimal possibilities for the dark world. Such exploration
will also help us fully understand the implications of data from
current and future dark matter experiments. One such example is the
recently proposed ``partially interacting dark matter'' scenario
(PIDM), in which only a subdominant component of dark matter undergoes
significant self-interactions~\cite{Fan:2013yva, Fan:2013tia}. In one
subclass of this scenario, ``double-disk dark matter'' (DDDM), this
subdominant component is charged under a long-range force which leads
to dissipative dynamics and the formation of a dark disk.\footnote{There
  is an argument that all dark matter could be
  dissipative~\cite{Foot:2013vna}.}

This scenario opens up interesting new possibilities for the dark
world.  Partially interacting dark sectors are less constrained by
halo-shape observations~\cite{Dave:2000ar, Rocha:2012jg, Peter:2012jh,
  MiraldaEscude:2000qt, Buote:2002wd}, relative to the early works on
self-interacting dark matter, which always assumed that all dark
matter self-interacts~\cite{Goldberg:1986nk, Khlopov:1989fj,
  Berezhiani:1995am, Spergel:1999mh, Mohapatra:2000qx,
  Mohapatra:2001sx, Foot:2004pa, Higaki:2013vuv, Feng:2008mu,
  Ackerman:2008gi, Feng:2009mn, Kaplan:2009de, Cline:2012is, CyrRacine:2012fz}.
However meaningful constrains on PDDM apply from dark acoustic oscillations (DAO)~\cite{Cyr-Racine:2013fsa}.
On the other hand, the partially dissipative scenario makes several
distinctive predictions for dark matter spatial and velocity
distributions as well as composition, yielding dramatical novel
predictions for both direct and indirect detection signals, as we now
detail. 

First, the spatial distribution of partially dissipative DM differs
significantly from the usual ellipsoidal dark matter halos.  While the
dominant component of dark matter, e.g., axions or some neutral
component of the dark sector, still acts as collisionless cold dark
matter and forms halos, the subdominant self-interacting dark matter
could (partially) cool via its long-range self-interaction and
collapse into a smaller volume within the larger halo, analogous to
the collapse of baryons into a disk through electromagnetic
interactions.  If the cooling is sufficiently rapid, a dark disk could
be formed, as happens in the DDDM scenario.  The collapsed spatial profile of partially or wholly
cooled dissipative dark matter results in
very different predictions for dark matter abundance, both locally in
our solar system and in galaxy halos, affecting both the types and
magnitudes of signals for {\it any} kind of dissipative dark matter detection.
The term ``dark disk'' has already been employed in the literature in
reference to the possibility that dark matter accretes onto the
baryonic disk during mergers~\cite{Read:2008fh}.  However, besides the
 differing origin of our dark disk, dissipatively-formed
dark disks have an important difference which leads to sharply
different observational consequences.  In the gravitationally formed
``dark disk'' scenario, dark matter in the disk is the same
particle species as the dark matter in the halo, while in the DDDM
scenario, the subdominant dark matter mostly resides in the disk and
has different properties and interactions than the dark matter forming
the bulk of the halo.

Second, the velocity distribution of a cooled sub-dominant component
of dark matter will have a smaller thermal velocity dispersion
compared to that of standard cold dark matter. Also, due to the
altered spatial distribution of dissipative dark matter, its velocity
distribution will in general be far from isotropic, and the
relative velocity between this component of dark matter and the solar
system can be very different from standard expectations. In the usual
cold dark matter scenario, the average velocity of dark matter with
respect to the Sun is of order the Sun's rotational velocity. In the
DDDM scenario, if the dark disk co-rotates with the baryonic disk, the
relative velocity of DDDM to the Sun is instead of order the Sun's
peculiar velocity, an order of magnitude below the Sun's rotational
velocity.

Third, a dissipative dark sector necessarily contains multiple
particle species.  For cooling to happen, a light species with
long-range interactions must exist, requiring the abundance of the
light species to be set by a number asymmetry.  This also implies that
an additional (``heavy'') particle species with equal and opposite
charge under the long-range dark force must exist with an equal number
asymmetry.  Generically, as we will review in Sec.~\ref{sec:review},
both an asymmetric and a symmetric relic abundance of the heavy
species will co-exist.

In this paper, we will show that due to these new characteristic
features of partially dissipative dark matter, dark matter direct 
detection constraints
on cool dark matter in PIDM/DDDM scenarios will be considerably
relaxed compared to those of normal cold dark matter, and in some
cases can even disappear.  To understand constraints coming from indirect detection, we focus on
possible signals from solar capture. Since the velocity dispersion of 
cooled DM is smaller than that of standard cold DM, cooled DM is more easily captured in massive bodies.  When there is a symmetric component of the relic
abundance, dark matter
annihilations inside the Sun yields signals in neutrino telescopes; when the asymmetric population
dominates, dark matter instead builds up inside the Sun, yielding potential
constraints from helioseismology~\cite{Spergel:1984re,Basu:2009mi}. 

Depending on the cooling time and other parameters of the dark sector (reviewed in
Section~\ref{sec:review}), the present-day spatial and velocity
distributions of the dissipative DM can vary. Therefore in this study
we will be agnostic both about the exact values of the velocity
dispersion and the local number density of the PIDM, keeping them as
free parameters of our theory, without necessarily assuming full
cooling into a dark disk 
distribution.  We will also not discuss here possible cosmic ray
signals of dissipative DM outside the solar system, 
since these observables are very sensitive to our assumptions about
the spatial distribution of the PIDM.

We also emphasize that many of our results regarding direct and indirect measurements 
largely apply also to cold flows of regular cold non-interacting DM. Such flows
are predicted, for example, by models of late infall of non-virialized DM into the 
the Galactic halo. These streams will typically have much smaller velocity 
dispersion than the virialized 
DM~\cite{Sikivie:1999jv,Natarajan:2005fh, Natarajan:2010jx}. 

The paper is organized as follows: first we review the basics of
dissipative dark matter in Sec.~\ref{sec:review}.  In Sec.~\ref{sec:direct_detection}, we discuss the relaxation of direct detection constraints on the couplings between dissipative dark matter and the SM particles. In Sec.~\ref{sec:solarcap}, we study the general theory of solar capture of dissipative dark matter and highlight the differences with respect to the standard scenario. In Sec.~\ref{sec:neutrino}, we discuss constraints from neutrino telescopes such as IceCube on dissipative dark matter annihilation inside the Sun, and comment on capture in the Earth. In Sec.~\ref{sec:helioseismology}, we discuss helioseismology constraints on purely asymmetric dissipative dark matter accumulated inside the Sun.  We conclude and discuss future directions in Sec.~\ref{sec:conclusion}. The demonstration of the validity of key assumptions in our analysis of solar capture are relegated to the appendix.


\section{Review of dissipative DM }
\label{sec:review}

In this section we review the necessary ingredients of dissipative
dark matter scenarios and the resulting dynamics. This review closely follows discussions in
Refs~\cite{Fan:2013yva,Fan:2013tia}.

\subsection{Spectrum and relic abundance of dissipative DM}

The dissipative dark matter sector we consider here is broadly
analogous to our baryonic sector. There are two species of particles,
a light species denoted by $C$ (for ``coolant'') with mass $m_C$,
analogous to the electron, and a heavy one denoted by $X$ with mass
$m_X$, analogous to the proton. Both $C$ and $X$ transform under an unbroken
dark gauge group associated with a long-range dark force. The simplest
possibility is that the gauge group is Abelian, a $U(1)_D$, with a
coupling strength $\alpha_D$.  The light ``dark electron'' $C$
particles annihilate away efficiently in the early Universe through $C
\bar{C} \to \gamma_D\gamma_D$~\cite{Fan:2013yva,Fan:2013tia}. Thus any
current population of $C$ particles must be asymmetric. Without loss
of generality, we will assume that only $\bar{C}$ exists now. To keep
the universe neutral under the $U(1)_D$, there must be a compensating
asymmetric number abundance of $X$ particles, which we take to have
equal and opposite charge as $\bar C$.

On the other hand, the annihilation rate of $X\bar{X} \to
\gamma_D\gamma_D$ is much slower than that of $C\bar{C}$ as the
annihilation rate decreases with increasing mass. Thus it is possible
to have a symmetric thermal relic abundance of $X, \bar{X}$ on top of
the asymmetric relic abundance of $X, \bar{C}$.
Therefore the most generic scenario for dissipative dark matter
includes both a symmetric and asymmetric component.  The relative
fraction of symmetric and asymmetric components in the relic density
has no impact on direct detection, in the experimentally interesting
case where the heavy species $X$ dominates the dark sector couplings
to the SM.\footnote{Cosmological constraints on the number of massless
  species in the early universe strongly constrain couplings between
  $C$ and the SM sector.} However, this fraction does affect the signals
resulting from capture of dissipative dark matter by stars. In
scenarios with a mixture of asymmetric and symmetric components, one
might expect high energy neutrino signals from captured dark matter
annihilating into SM particles, as we will study  in
Sec.~\ref{sec:neutrino}. If the dissipative dark matter is purely
asymmetric, there is no dark matter annihilation and thus no signal
for neutrino telescopes. But since in this case stars will accumulate dark matter
 without any annihilation to reduce dark number density, the
helioseismology constraints that we will discuss in
Sec.~\ref{sec:helioseismology} could  potentially become important.

\subsection{Dissipative dynamics}

In the early Universe, the asymmetric component of the relic abundance, made of $X$ and
$\bar{C}$ particles, largely recombines into dark atoms once the
temperature of the universe drops below the binding energy. After the
dark matter particles fall into the galactic halo, however, they will
be shock heated to the halo virial temperature, which is generally
high enough to ionize the dark atoms and form a fully ionized dark
plasma.  A symmetric relic density of $X$, $\bar X$ adds additional
dark ions to the plasma.

The existence of a long-range dark force then allows the dark plasma
to cool through:
\begin{itemize}
\item{Bremsstrahlung process: $X \bar C \to X\bar C\gamma_D$. Dissipative dark
    matter scatters, emitting a soft dark photon which carries away
    energy;}
\item{Compton scattering: $\bar C(X) \gamma_D \to \bar C(X) \gamma_D$. Dark
    matter particles scatter off dark photons, depositing energy into
    the dark CMB.}
\end{itemize}
When the dark plasma is cooled enough that dark recombination can
happen again, further cooling through atomic or molecular processes
will take place, which we do not consider further. 

Since the rates of both bremsstrahlung and Compton
scattering increase with decreasing mass, it is dominantly the light
particle $\bar C$ which can efficiently transfer energy to the dark
radiation.  One necessary condition for sufficient cooling to form a
dark disk is that the cooling time scale is shorter than the age of
the Universe. This requirement sets a constraint on the mass of the
``coolant'' particle, $m_C \lesssim 1$ MeV.  The Rutherford scattering
between $X$ and $\bar C$ particles transfers energy between $X$ and $\bar C$,
allowing the entire plasma to cool. In the portion of parameter space
in which the time scale for energy equipartition is shorter than the
cooling time scale, cooling proceeds adiabatically. For very small
$m_C/m_X$, the Rutherford scattering rate is so slow that the plasma
may cool out of equilibrium.

Similar to baryons, the dark plasma acquires angular momentum via
tidal torques during structure formation. Thus
when both the cooling time scale and energy equipartition time scale
are shorter than the age of the Universe, the dark plasma could
potentially form a rotationally supported disk.  However, it is also
of interest to consider a broader range of parameter space, in which
either the cooling or the energy equipartition time scale is longer
than the age of the Universe.  Here some partial cooling will still
occur and the dark plasma will collapse into a smaller volume in the
halo without forming a disk. This nonequilibrium regime could be
fruitfully addressed by future $N$-body simulations.

We caution the reader that a very thin disk might be unstable in the sense that it may fragment and large gaseous clouds may form. We estimated the Jeans mass for this fragmentation in a previous paper~\cite{Fan:2013yva}. Once clouds above the Jeans mass begin to collapse, further cooling through atomic and molecular processes might lead to the formation of dark stars. In the absence of numerical simulations it is difficult to further quantify these effects.

\subsection{Velocity distribution and density distribution}
\label{sec:vel}
The velocity dispersion of dissipative dark matter at the present day, $\overline{v}\equiv \sqrt{\langle v^2\rangle}$, is set by the temperature at which cooling stops, $T_{\rm {cooled}}$. As a crude estimate, we expect that cooling stops when dark recombination happens. 
The velocity dispersion is then estimated to be
\beq
\overline{v}& \approx &\sqrt{3 \frac{T_{\rm {cooled}}}{m_X} }\nonumber  \\
&=&\sqrt{3 \frac{r B_{XC}}{m_X} }=\sqrt{ \frac{3 r}{2} \frac{\alpha^2 m_C}{m_X} }\nonumber \\
&=& 10^{-4} \frac{\alpha}{10^{-2}} \sqrt{\frac{r}{0.1} \frac{m_C}{1\,{\rm MeV}}\frac{1\,{\rm GeV}}{m_X}},
\label{eq:velocity}
\eeq 
where $r \equiv T_{\rm {cooled}}/B_{XC}$ with $B_{XC}$ the binding energy of the dark atom. A rough estimate using the Saha equation shows that $r$ is in the range $(0.02 - 0.2)$~\cite{Fan:2013yva,Fan:2013tia}.

As one can see from Eq.~\ref{eq:velocity}, the thermal velocity dispersion of dissipative dark matter could be much smaller than that of normal cold dark matter, which is $\mathcal{O}(10^{-3} c)$. Strictly speaking, this estimate for the velocity dispersion should only be trusted when sufficient cooling happens.  If the cooling time scale is slightly longer than the age of the universe, the temperature of the dark plasma nowadays will generally be above the binding energy. Even in the case when sufficient cooling occurs, we neglect additional heating and cooling from atomic and molecular processes, which have opposite effects on the velocity dispersion. Thus the estimate of Eq.~\ref{eq:velocity} should be taken with a grain of salt. More robust estimates would require future numerical simulations. For our purpose, it suffices to take the velocity dispersion as a free parameter, varying from $(10^{-4} - 10^{-3})c$,\footnote{Even smaller velocity dispersions, $\bar{v} \ll 10^{-4}$, may be possible. But such small dispersions will not affect the results of our analysis for either direct detection or solar capture as both 
direct and indirect detection signals become insensitive to $\bar v$
when it is much smaller than the relative velocity between the solar system and the dark matter.} and study the resulting impact on direct and indirect detection.

Another important velocity relevant for both direct and indirect detection signals is the relative velocity between the flux of dark matter particles and the 
Sun. If dissipative dark matter particles cool into a rotationally supported disk as in the DDDM scenario, they will move around the 
center of the galaxy in largely coplanar circular orbits. If the dark disk is aligned and co-rotating with the baryonic disk, then in the 
vicinity of the Sun both dark matter particles and baryonic structures will move in the same mean circular orbit. This means that the average relative 
velocity of dark matter particles with respect to the Sun comes from deviations from the baryonic disk's average rotational velocity, that is the 
peculiar velocity of the Sun, of order $|v_{\rm rel}^\odot| \sim 10^{-4}$. Similarly the relative velocity between the flux of dark matter particles and 
the Earth is the Earth's peculiar velocity, which is again of order $|v_{\rm rel}^\oplus| \sim 10^{-4}$. These observations also hold true for normal 
cold dark matter in a dark disk formed by accretion onto the baryonic disk, as first noticed in~\cite{Bruch:2008rx, Bruch:2009rp}. However, there 
is one major difference between our dissipative dark disk and their accretional dark disk. In Refs.~\cite{Bruch:2008rx, Bruch:2009rp}, cold dark matter has both a halo component and a disk component while in our scenario, dissipative dark matter particles mostly reside in the disk. This will lead to a relaxation of the constraints on the couplings between dark matter particles and the SM particles from direct detections, which will be demonstrated in the following section. If the dark disk anti-rotates with the baryonic disk, or if the dissipative dark matter only cools into a non-rotational clump, the relative velocities are still of order ${\cal O} (10^{-3})$, comparable to that of ordinary cold dark matter. Thus in what follows we focus on relative velocities in the physically interesting range $(10^{-4} - 10^{-3})c$. 

In our analysis, we assume that the dissipative dark matter has a Maxwell-Boltzmann velocity distribution. For both direct and indirect searches which we will discuss, the signal rate is only sensitive to ${\rm max} (|v_{\rm rel}|, \bar{v}$) when these two velocities are of different orders of magnitude. In Table~\ref{table:velocities}, we list all velocities relevant to direct and indirect detection signals in order to fix the notation we will use below. 

\begin{table}
\begin{center}
\begin{tabular}{|c|c|c|}
\hline 
$\vec{v}_{\rm circ}$ & Sun's rotational velocity & (0, 220, 0) km/s \\
\hline
$\vec{v}_\odot$ & Sun's peculiar velocity & (10, 5.25, 7.17) km/s \\
\hline
$\vec{v}_\oplus$ & Earth's peculiar velocity & $\vec{v}_\odot +$29.8 $ f(t)$ km/s\\
\hline
$v_{\rm gesc}$ & escape velocity of the galaxy & 540 km/s \\
\hline
$v_{\rm esc}^{\odot}(r)$ & escape velocity inside the Sun as a function of radius & $v_{\rm esc}^{\odot}(0) = 1386$\, km/s \\
&&$v_{\rm esc}^{\odot}(R_\odot) = 618$ km/s \\
\hline
$\bar{v}$ & DM velocity dispersion  & $10^{-4} - 10^{-3}$  \\
\hline
$\vec{v}_0$ & average velocity of DM in the galactic frame & ${\cal{O}}(10^{-3})$ \\
\hline
$\vec{v}_{\rm rel}^\odot$ & relative velocity of dissipative DM to the Sun & $\vec{v}_0 - \vec{v}_{\rm circ} -\vec{v}_\odot \subset (10^{-4} - 10^{-3}) c$\\
\hline
\hline
\end{tabular}
\caption{Velocities relevant to solar capture and direct detection. We use coordinates where $x$ points towards the center of the galaxy, $y$ in the direction of the disk rotation, and $z$ towards the galactic north pole. For the Earth's peculiar velocity, the time dependence is given by $f(t) = \cos(2\pi(t-t_{\rm June}))(0.262,0.504,-0.823)+\sin(2\pi(t-t_{\rm June})) (-0.960, 0.051,-0.275)$ with $t_{\rm June} =$ June 2nd. The solar escape velocities are calculated from standard solar model with parameters in~\cite{Bahcall:2004pz}. }
\label{table:velocities}
\end{center}
\end{table}

Finally we comment on the local density of dissipative dark matter near the Sun. This quantity depends crucially on the  dark disk thickness and the alignment of the dark disk with respect to the baryonic disk.  In general, there is only a weak constraint on the local dissipative dark matter density from the Oort limit, which allows a local density up to a few GeV/cm$^3$~\cite{Fan:2013yva}. The Oort limit is derived from subtracting from the measured overall surface density below a height $z_0$ in the Milky Way the contribution from the stellar disk, gas disk and other visible baryonic matter.\footnote{The most recent  analysis of surface density can be found in~\cite{Bovy:2013raa}, though it does not quote error bars for some contributions from 
baryonic matter, e.g., the contribution from the interstellar gas.} Another type of analysis based on a hypothesis that impact craters form at an enhanced rate when the Sun passes through the dark disk hints that the DDDM local density could vary from zero to a few GeV/cm$^3$~\cite{Matt:crater}. Thus in our analysis, we will take the local density of dissipative dark matter to be a free parameter in the range (0 - a few) GeV/cm$^3$. For a fully cooled dark disk, this range of values for the local
dark-matter density is estimated to allow a stable disk within the
approximations of~\cite{Fan:2013yva, Fan:2013tia}.

Strictly speaking, the local densities relevant to direct detection and solar capture are different. We will consider the most optimistic case for both signals, where a dark disk is precisely aligned with our baryonic disk. As the Sun could oscillate around the plane of dark disk, for direct detection experiments, the relevant local density is the \emph{current} density of dissipative dark matter in the solar system. On the other hand,  for indirect detection depending on solar capture, it is the local density \emph{averaged} over the Sun's age that matters. As long as the oscillation period is much shorter than the Sun's age, the averaged local density is only sensitive to the local density in the middle of the dark plane. In our analysis, we do not differentiate between the current and the averaged local densities. But it is worthwhile to emphasize that although currently the solar system could be outside the dark disk and thus direct detection experiments could be completely insensitive to the dark disk, indirect detection could still potentially set interesting constraints.

\section{Direct detection of dissipative DM}
\label{sec:direct_detection}

Direct detection of dissipative dark matter is highly model dependent and can be evaded easily. For instance, in the DDDM scenario, if the dark disk is not aligned with the baryonic disk, or if the dar disk is aligned with the baryonic disk but is so thin ($\lesssim 10$ pc) that the Sun is outside the DDDM disk, there will be no direct detection signals at all. On the other hand, if the two disks are approximately aligned and the solar system is inside the DDDM disk,  predictions for direct detection can be dramatically different from that of normal cold dark matter. In this section, we will assume the most optimistic case for direct detection, with our solar system inside the DDDM disk, and focus on the possibility of dissipative dark matter scattering elastically off target nucleons in detectors. We will show that in this case, the direct detection constraints can be considerably relaxed. The possibility of inelastic scattering of light dark matter as an explanation for the three possible signal events observed in CDMS II silicon data~\cite{Agnese:2013rvf} was discussed in~\cite{McCullough:2013jma}.

\begin{figure}[!h]\begin{center}
\includegraphics[width=0.4\textwidth]{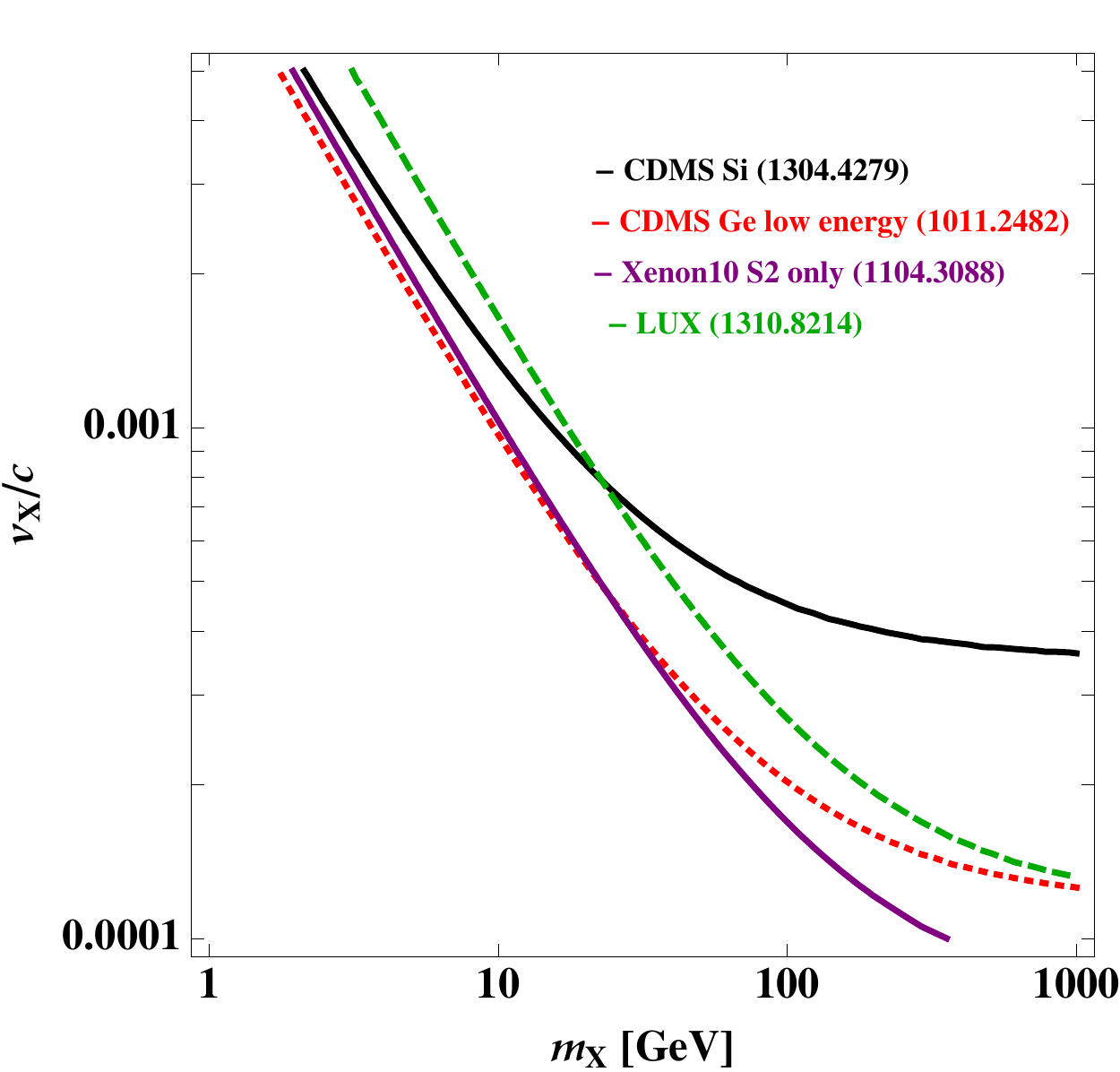}
\end{center}
\caption{The parameter space in the ($m_X, v_X/c$) plane to which different direct detection experiments are sensitive; the region of sensitivity is above each curve.}
\label{fig:directdetectionregion}
\end{figure}

First we briefly review the kinematics of elastic scattering of dark matter particles in direct detection. A dark matter particle moves with a nonrelativistic velocity, $v_X$, in the lab frame, then scatters off a nucleus in the detector. Depending on the scattering angle, the recoil energy imparted to the nucleus varies from zero to
\beq
E_R^{\rm max}& = & \frac{2 \mu_N^2}{m_N} v_X^ 2 \label{eq:recoil}\\
& \approx & 0.5\, {\rm keVnr}\left( \frac{\mu_N}{50\,{\rm GeV}} \right)^2\frac{100\,{\rm GeV}}{m_N}\left(\frac{v_X}{10^{-4}}\right)^2,\nonumber
\eeq
where $m_N$ is the mass of the target atom, and $\mu_N$ is the reduced mass of the dark matter--nucleus system.  Most experiments are only sensitive to energies above a threshold energy, $E_R^{\rm thr}$, below which noise and different backgrounds overwhelm possible dark matter signals.  The typical threshold for nuclear recoil energies in current direct detection experiments is a few keV. Having a threshold $E_R^{\rm thr}$ means that for a given dark matter particle mass $m_X$ each experiment is only sensitive to a minimum value of the dark matter relative velocity $v_X^{\rm min}$:
\beq
v_X^{\rm min}= \sqrt{\frac{E_R^{\rm thr}m_N}{2\mu_N^2}}. 
\eeq
For elastic scattering of heavy dark matter with a given DM mass of order ${\cal O} (100$ GeV), choosing heavier nuclei reduces the DM velocity threshold $v_X^{\rm min}$. Thus direct detection detectors with heavier nuclei will sample more of the dark matter velocity distribution and have greater sensitivity. This is shown in Fig.~\ref{fig:directdetectionregion}, in which we plot regions in the ($m_X,  v_X/c$) plane to which each of some representative 
direct detection analysis~\cite{Ahmed:2010wy, Agnese:2013rvf, Angle:2011th, Akerib:2013tjd} is sensitive. It is clear from the figure that for a heavy dark matter particle with mass around 100 GeV, the relative velocity has to be about or above $10^{-4}$ to trigger a signal in at least one of the direct detection experiments.

Now consider a dark matter flux with a velocity distribution $f(\vec{v})$. The rate for spin-independent elastic scattering is:
\beq
\frac{d R}{d E_R}=N_t \frac{m_N \rho_X \sigma_n}{2 m_X \mu_{n}^2} A^2 F(E_R)^2{E_R}\int_{v_{\rm min}}^{v_{\rm gesc}} d^3v \frac{f(\vec{v})}{|\vec{v}|},
\eeq
where $N_t,~m_N$ and $A$ are the number, mass, and atomic number of the target atoms; $m_X$, $\rho_X$ and $\vec{v}$ are the the mass, local density, and velocity of DDDM at the Sun; $\sigma_n$ is the zero-momentum spin-independent DDDM--nucleon scattering cross section; $\mu_{n}$ is the reduced mass of the DDDM--{\em nucleon} system; $F(E_R)^2$ is the nuclear form factor; $v_{\rm min}$ the minimum DDDM velocity needed to create a nuclear recoil with recoil energy $E_R$ and $v_{\rm gesc}$ the galactic escape velocity of DDDM. We assume that the DDDM couplings to all nucleons are equal for simplicity. Before taking into account the nuclear form factor $F(E_R)^2$, the spectrum is flat between 0 and $E_R^{\rm max}$. However, the nuclear form factor $F^2(E_R)$ is in general an exponentially falling function, which suppresses higher energy recoils, yielding a falling spectrum with an end point at $E_R^{\rm max}$.  Thus the shape of the recoil spectrum for elastic scattering of a DDDM particle off nucleons is still similar to that of ordinary cold dark 
matter.\footnote{The DDDM recoil spectrum is steeper than that of normal dark matter but given the exponentially falling backgrounds, it is challenging to resolve the different spectra in realistic experimental settings.}

\begin{figure}[!h]\begin{center}
\includegraphics[width=0.5\textwidth]{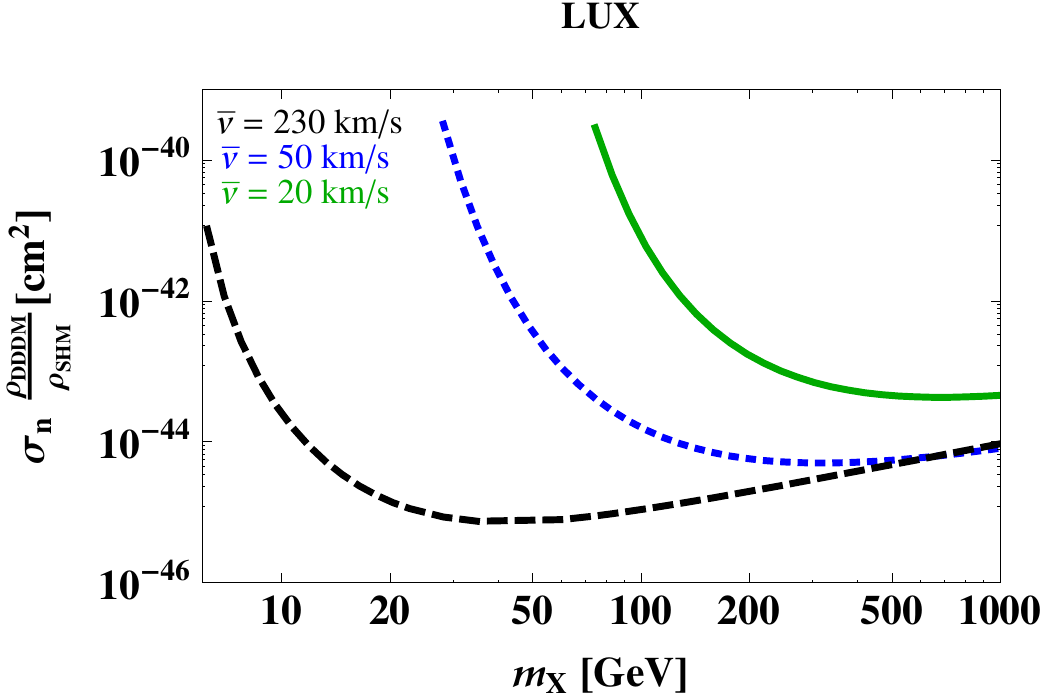}
\end{center}
\caption{Bounds from the LUX experiment in the ($m_X, \sigma_n \rho_{DDDM}/\rho_{\rm {SHM}}$) plane for different velocity dispersions. Black dashed: ordinary cold dark matter distribution with $\bar{v} = 230$ km/s; blue dotted: $\bar{v} = 50$ km/s; green solid: $\bar{v} = 20$ km/s.  }
\label{fig:directdetectionbound}
\end{figure}

We assume that the dissipative dark matter velocity distribution is given by a Maxwell-Boltzman distribution in the frame of the detector
\beq
f(\vec{v}) = \frac{1}{2\pi^{3/2} \bar{v}^3} e^{-\frac{\left|\vec{v} - \vec{v}_{\rm rel}^\oplus\right|^2}{\bar{v}^2}},
\eeq
with $\bar{v}, \vec{v}_{\rm rel}^\oplus$ corresponding to the velocity dispersion and relative velocity with respect to the Earth respectively.
The strongest current constraints of all the direct detection experiments come from LUX~\cite{Akerib:2013tjd}, and we plot them in 
Fig.~\ref{fig:directdetectionbound}.
We assume the average velocity of DDDM in the galactic frame, $\vec{v}_0$, is the rotational velocity of the baryonic disk and thus
 $ \vec{v}_{\rm rel}^\oplus = \vec{v}_\oplus$, the peculiar velocity of the Earth. We plot the constraints on the DM-nucleon cross section, $\sigma_n$, times the ratio of the local DDDM density normalized by the normal cold dark matter density near the Sun, $\rho_{\rm SHM}$ = 0.4 GeV/cm$^3$, for two different velocity dispersions $\bar{v} = 50$ km/s and $\bar{v} = 20$ km/s. We used Yellin's maximal gap method~\cite{Yellin:2002xd} to set limits. We also plot the constraints on normal cold dark matter with standard halo model, $\bar{v} = 230$ km/s, for comparison. For velocity dispersions smaller than the peculiar velocity of the Earth, however, the constraints will not be relaxed further as the direct detection is only sensitive to ${\rm max} (\bar{v}, |v_{\rm rel}^\oplus|)$.

In Fig.~\ref{fig:directdetectionbound} we see that, if the dissipative dark matter velocity dispersion is as small as the relative velocity $\bar{v} \sim |v_{\rm rel}^\oplus|  \lesssim 10^{-4}c$, for dark matter with mass about or below 70 GeV, a big scattering cross section scattering off nucleons, $\sigma_n\sim 10^{-39}$ cm$^2$, which is of order of $Z$-exchange cross section, is still allowed assuming the local density of dissipative dark matter is the same as that of normal cold dark matter at the Sun, 0.4 GeV/cm$^3$. If the local density of the dissipative dark matter near the Sun is smaller than 0.4 GeV/cm$^3$, 
allowed values for $\sigma_n$ can be even larger. Even if the local density near the Sun is one order of magnitude above 0.4 GeV/cm$^3$, for 
70~GeV DDDM with velocity dispersion of $10^{-4}c$, the allowed scattering cross section is $\sigma_n\sim 10^{-40}$ cm$^2$, namely, five orders of magnitude larger than that permitted for normal cold dark matter.   

In summary, due to the small velocity dispersion of DDDM, only the energy bins close to an experimental threshold are sensitive to DDDM scattering. The constraints on the cross sections for DDDM scattering off nucleons are greatly relaxed, and for velocity dispersion $\lesssim 10^{-4} c$, a large cross section of order the $Z$-exchange cross section is still allowed for DM with mass below 70~GeV!  So far, the importance of understanding and improving energy calibration around the threshold has been mostly emphasized for ruling in or out the light DM scenario. Yet from the discussions above, pushing direct direction thresholds  lower could also be important for the DDDM scenario, or in general, for the detection of any  dark matter component with a low mean velocity.

\section{Solar capture of dissipative DM }
\label{sec:solarcap}
We demonstrated in the previous section that when dissipative DM has both small velocity dispersion and small mean velocity relative to the Earth, direct detection experiments set much weaker constraints on the coupling between dissipative dark matter and the SM than is the case for normal cold dark matter.
Now we turn to the indirect constraints on this dissipative dark matter scenario coming from the solar capture of dark matter.   In this section, 
we will first derive the self-capture rate for dissipative dark matter within the Sun. Due to the long-range dark force, the scattering cross section 
for self-capture is velocity-dependent. This results in an enhanced self-capture rate with a very different parametric dependence from the one derived 
in Ref~\cite{Zentner:2009is}, which assumed that the differential cross section for the capturing process is velocity-independent. We then discuss two
limiting cases: capture of dark matter with a symmetric relic abundance and capture of dark matter with an asymmetric relic abundance. In general, 
dissipative dark matter will be a mixture of both symmetric and asymmetric relic abundances and thus its capture interpolates between the two limits. 
We end this section by writing down the most general equations governing the capture of dissipative dark matter particles.

\subsection{Self-capture of dark matter particles }
\label{sec:selfcap} 
We begin the discussion of solar capture by deriving the self-capture rate per target captured DM particle. We follow the standard procedure for calculating the capture rate as developed by Gould in Refs~\cite{Gould:1987ww, Gould:1987ir, Gould:1991hx}. The main novelty in the derivation is that the differential cross section of the self-capture process is now velocity-dependent.\footnote{For capture by nucleons, the nuclear form factor will also introduce a weak velocity dependence to the cross section.} This will lead to a dramatically different parametric dependence of the final result. 

In Gould's procedure, one first considers capture in an individual spherical shell of a massive body, here the Sun, of radius $r$ 
and local escape velocity $v_{\rm esc}(r)$.
For the Sun, $v_{\rm esc}^\odot(0) \approx 1386$ km/s and $v_{\rm esc}^\odot(R_\odot) \approx 618$ km/s. Outside the shell, consider a bounding 
surface of radius $R$ with $R \gg r$ such that the gravitational field due to the Sun is negligible at $R$. The one-dimensional speed distribution function of dissipative dark matter particles at $R$ is given by $f(u)$, with $u$ the speed at $R$,
\beq
f(u)=\sqrt{\frac{6}{\pi}}\frac{1}{\bar{v}}x^2 e^{-x^2}e^{-\eta^2}\frac{\sinh (2x\eta)}{x\eta}~,
\eeq
where the dimensionless variables are given by $x^2 = 3(u/\bar{v})^2/2$ and $\eta^2= 3 (v_{\rm rel}^\odot/\bar{v})^2/2$ with $v_{\rm rel}^\odot$ corresponds to the relative velocity between dark matter flux and the Sun. The infalling  dark matter particles reach the shell at $r$ with a 
speed $w = \sqrt{u^2 + {v_{\rm esc}^\odot(r)}^2}$. Taking $\Omega(w)$ to be the rate at which a dark matter 
particle with speed $w$ at the shell $r$ scatters to a speed less than $v_{\rm esc}^\odot(r)$ while the target particle does not gain energy 
above the local escape energy, the capture rate of $X$ particles per unit shell volume as an integral over the speed distribution at infinity is then given by 
\beq
\frac{d C_S}{dV} = \int \frac{f(u)\,n_X}{u\, N_t}\, w \,\Omega(w) du,
\eeq
where $N_t$ is the total number of captured dissipative dark matter particles inside the Sun and $n_X$ is the local number density of $X$. 

The rate $\Omega(w)$ is given by $n_t(r) \sigma_{\rm cap} w$, with $n_t(r)$  the number density of dark matter targets at the location of the shell and $\sigma_{\rm cap}$ the cross section of self-scattering processes leading to capture.  Under the assumption that  dark matter in the Sun can be described by a thermal distribution at the solar core temperature $T_\odot = 1.57 \times 10^7$ K, the target number density is $n_t(r) = n_t(0) {\rm exp}[-M_X \phi(r)/T_\odot]$, where $\phi(r)$ is the gravitational potential as a function of position within the Sun. We will demonstrate the validity of the thermalization assumption in the Appendix.
In the dissipative dark matter scenario, the dominant self-capture process, e.g., $X$ ions captured by bound $X$ (or $\bar{X}$) ions, proceeds through Rutherford scattering with the differential cross section 
\beq
\label{eq:Ruth}
\frac{d\sigma_R }{d \Omega} = \frac{\alpha_D^2}{4m_X^2 w^4 \sin^4 \left( \frac{\theta}{2} \right)}~.
\eeq  
Again, in adopting the formula above we assumed that the target dark matter particles are in thermal equilibrium with the solar core. The velocity $w$ of the infalling dark matter particles is much larger than the thermal velocity of the target dark matter particles and thus we can always approximate their relative velocity by $w$. The infrared divergence $\theta \to 0$ is regulated by the fact that there is always a minimal scattering angle associated with the capture: an incoming dark matter particle has to lose at least a fraction $u^2/(u^2+v_{\rm esc}^\odot(r)^2)$ of its kinetic energy to be captured. Also taking into account that the target dark matter particle cannot gain energy at or above $m_X v_{\rm esc}^\odot(r)^2/2$ (otherwise, there will be no net gain in the number of captured dark matter particles), for net capture to occur the cosine of the scattering angle must lie in the range 
\beq
\label{eq:thetacap}
 \frac{u^2-v_{\rm esc}^\odot(r)^2 }{w^2}\leq\cos \theta \leq \frac{v_{\rm esc}^\odot(r)^2 - u^2}{w^2}\; ;
\eeq
a necessary but not sufficient condition for capture to take place is $u < v_{\rm esc}^\odot (r)$. 
For cooled dark matter, the average velocity $\langle u \rangle$ is much smaller than the escape velocity
inside the Sun, $\langle u \rangle = \bar{v} \ll v_{\rm esc}^\odot$, and this condition is almost always satisfied. 
The cross section for capture without ejection is then 
\beq
\sigma_{\rm cap} =  \frac{\pi\alpha_D^2}{m_X^2 w^2u^2}\left(1-\frac{u^2}{v_{\rm esc}^\odot(r)^2}\right).
\eeq
Notice that it is enhanced by $w^2/u^2$ compared to the cross section for hard Rutherford scattering with order one scattering angles, $\sigma_{\rm hard} \approx   \frac{\pi\alpha_D^2}{m_X^2 w^4}$. 

Another subtlety associated with capture due to Rutherford scattering is that there is a finite impact parameter above which the dark charges of either the incoming dark matter particles or the dark matter targets are screened, and for impact parameters larger than this charge-screening length, the interaction is suppressed. For the capture of dark ions, this length is the Debye screening length of the dark plasma inside the Sun,
\beq
\lambda_D=\sqrt{\frac{T_\odot}{4\pi\alpha_D\,n_t(0)}}\approx 0.2\, {\rm km}\sqrt{\frac{10^{-2}}{\alpha_D} \frac{2 \,{\rm cm}^{-3}}{n_t(0)}},
\eeq
where $n_t(0)$ is the captured dark matter density at the center of the Sun. Notice that $n_t(0)$ is \emph{time-dependent}, and $\lambda_D$ decreases with time as the number of accumulated dark matter targets increases. 
For the capture of dark atoms, the charge-screening length is the Bohr radius of the dark atom
\beq
r_B = \frac{1}{\alpha_D m_C} = 2 \times 10^{-9} \,{\rm  cm} \,\frac{10^{-2}}{\alpha_D} \frac{1\, {\rm MeV}}{m_C}.
\label{eq:Bohr}
\eeq
At scales smaller than $r_B$, the dark atom capture is dominated by the capture of the $X$ nucleus inside the dark atom through Rutherford scattering. Above $r_B$, the capture of dark atom has to go through short-range dipole interactions, for which the cross sections are small. 
The maximal impact parameter is related to the the minimal velocity $u_{\rm min}$ through
\beq
b_{\rm max} = \sqrt{\frac{\sigma_{\rm cap}^{\rm max}}{\pi}}= \frac{\alpha_D}{m_X\,w\,u_{\rm min}}\sqrt{1-\frac{u_{\rm min}^2}{v_{\rm esc}^\odot(r)^2}}.
\eeq
Thus for Rutherford capture to be effective, the minimal velocities for dark ions and dark atoms have to be 
\beq \label{eq:vcutoff}
u_{\rm min}&\approx&2 \times 10^{-20}\left(\frac{\alpha_D}{10^{-2}}\right)^{3/2}\frac{100\,{\rm GeV}}{m_X}\sqrt{\frac{n_t(0)}{2\,{\rm cm}^{-3}}}, \quad {\rm dark\, ion}\\
u_{\rm min} &\approx&2 \times 10^{-7}\left( \frac{\alpha_D}{10^{-2}}\right)^2\frac{ m_C}{1\,{\rm Mev}}\frac{100\,{\rm GeV}}{m_X} \quad {\rm dark\, atom}.
\eeq

Plugging in for the cross section and velocity distribution function and integrating over the velocity $u$ and volume $V$, the self-capture rate is given by 
\beq 
C_S&=& \frac{\pi \alpha_D^2n_X}{m_X^2N_t} \int n_t(r)\ dV \int_{u_{\rm min}}^{v_{\rm gesc}} du\frac{f(u)}{u^3}\left(1-\frac{u^2}{{v_{\rm esc}^\odot(r)}^2}\right) \nonumber \\
&\approx& \frac{3 \sqrt{6\pi}n_X}{\bar{v}^3} \frac{\alpha_D^2}{m_X^2} e^{-\eta^2} \log \left(\sqrt{\frac{2}{3}}\frac{\bar{v}}{u_{\rm min}}\right) \quad \eta \lesssim 1 \\
&\approx & \frac{\pi n_X}{|v_{\rm rel}^\odot|^{ 3}}\frac{\alpha_D^2}{m_X^2} \quad \eta \gg 1,
\eeq
where in the first line the upper limit of integration is the galactic escape velocity $v_{\rm gesc}$, which is always smaller than the Sun's escape velocity $v_{\rm esc}^\odot(r)$. In the second and third line, we approximated $\left(1-\frac{u^2}{{v_{\rm esc}^\odot(r)}^2}\right)$ by 1 and take the two limits $\eta \lesssim 1$ and $\eta \gg 1$ (i.e, $\eta \geq 10$) to demonstrate the parametric dependence of the self-capture rate. We have checked numerically that the approximated formula agrees with the exact results up to the 5\% level. These capture rates are very different from the self-capture rate for a velocity-independent differential cross section discussed in~\cite{Zentner:2009is}. In that case, the velocity dependence of self-capture is given by $\frac{(v_{\rm esc}^\odot (R_\odot))^2}{\bar{v}}\frac{{\rm Erf}(\eta)}{\eta}$, which is approximately $\frac{(v_{\rm esc}^\odot (R_\odot))^2}{\bar{v}}$ for $\eta \lesssim 1$ and $\frac{(v_{\rm esc}^\odot (R_\odot))^2}{|v_{\rm rel}^\odot|}$ for $\eta \gg 1$. 

When $\eta \lesssim 1$, or equivalently when the relative velocity $|v_{\rm rel}^\odot|$ is comparable to the velocity dispersion $\bar{v}$, the main contribution to the integral is from small $u$ for which the integrand is approximately proportional to $\int du/u \, e^{-\eta^2}$. Thus the self-capture rate is logarithmically sensitive to $u_{\rm min}$ and therefore the charge-screening length. For capture of dark ions, as the Debye length is a time-dependent quantity, the self-capture rate of dark ions for $\eta \lesssim 1$ decreases logarithmically with time.

For $\eta \gg 1$, the relative velocity is much larger than the velocity dispersion. In this case, the contribution from small $u$ is suppressed by $e^{-\eta^2}$ and the integrand is maximized when $x^2 = 3(u/\bar{v})^2/2 \approx \eta^2$ or equivalently $u \approx |v_{\rm rel}^\odot|$. The self-capture rate is only sensitive to the relative velocity $|v_{\rm rel}^\odot|$ and is not enhanced for small velocity dispersion $\bar{v}$. 

The approximated analytic formulas confirm that only ${\rm max} (\bar{v}, v_{\rm rel}^\odot)$ matters for the self-capture rate. 
Given that the magnitude of the relative velocity cannot be smaller than that of the Sun's peculiar velocity ${\cal O} (10^{-4})$ and is always in the range of $\mathcal{O}(10^{-4} - 10^{-3}) c$, for $\eta \lesssim 1$, the velocity dispersion $\bar{v}$ cannot be smaller than ${\cal O}(10^{-4})$. For smaller velocity dispersions, $\bar{v} < 10^{-4}$, $\eta \gg 1$ is usually satisfied. Thus for a given relative velocity, velocity dispersions smaller than that relative velocity do not enhance the self-capture rate.

An upper bound for the total dissipative dark matter self-capture rate arises when the sum of the self-interaction cross sections over all dark matter targets is equal to the surface of the volume occupied by the targets: 
\beq
\langle \sigma_{\rm cap} \rangle N_t^* =\pi r_X^2,
\eeq
where $\langle \sigma_{\rm cap} \rangle$ is the capture cross section averaged with velocity distribution $f(u)$,
\beq \label{eq:avexseccap}
\langle \sigma_{\rm cap} \rangle \approx 10^{-24} \, {\rm cm}^2 \left( \frac{\alpha_D}{10^{-2}}\right)^2 \left(\frac{100\,{\rm GeV}}{m_X}\right)^2\left(\frac{10^{-3}}{\bar{v}}\right)^2 \frac{{ F(\eta)}}{\eta}, 
\eeq
where $F(\eta)$ is the Dawson integral $F(\eta) = e^{-\eta^2} \int_0^\eta e^{-y^2} dy$. When $\eta \ll 1$, $\eta/F(\eta) \approx 1$ and when $\eta \gtrsim 1$, $\eta/F(\eta) \approx 2 \eta^2$. 
The radius of the volume occupied by the targets, $r_X$, can be estimated as 
\beq\label{eq:rx}
r_X \approx\sqrt{ \frac{9}{4\pi} \frac{T_\odot}{G_N\rho_\odot m_X}} \approx 0.13 \sqrt{\frac{1\,{\rm GeV}}{m_X}} R_\odot,
\label{eq:scaleradius}
\eeq
where the Sun's radius is $R_\odot \approx 7\times 10^{10}$ cm. One can see that as long as captured dark matter particles are thermalized with the Sun's core, they will only occupy a small region inside the Sun. 

Once the geometric limit is satisfied, the self-capture rate becomes $C_S^{\rm eff}$, obtained via replacing $\sigma_{\rm cap} N_X$ in $C_SN_X$ by $\sigma_{\rm eff}$. Numerically, $C_S^{\rm eff}$ can be approximated as
\beq
C_S^{\rm eff} \approx 7 \times 10^{23}\, {\rm s}^{-1} \left(\frac{\rho_X}{0.4\,{\rm GeV}/{\rm cm}^3}\right)\left(\frac{100\,{\rm GeV}}{m_X}\right)^2\left(\frac{10^{-3}}{|v_{\rm rel}^\odot|}\right) {\rm Erf(\eta)}
\eeq
One can also estimate the total number of targets $N_t^*$ when the self-capture rate saturates the geometric bound,
\beq
N_t^*&=& \frac{\pi r_X^2}{\langle \sigma_{\rm cap} \rangle} \nonumber \\
&\approx& 2 \times 10^{-15}N_\odot \left(\frac{0.01}{\alpha_D}\right)^2\left( \frac{m_X}{100\,{\rm GeV}}\right) \left(\frac{\bar{v}}{10^{-3}}\right)^2 \frac{\eta}{F(\eta)},
\eeq
where $N_\odot \approx 10^{57}$ is the total number of nucleons in the Sun.

\subsection{Capture of purely symmetric self-interacting DM}
To make contact with the existing literature, we first consider the scenario with a negligible asymmetric relic abundance of $X, \bar{C}$. This scenario may not necessarily undergo (significant) cooling\footnote{The fraction of asymmetric abundance needed for cooling as a function of $m_C$ is shown by Fig.~6 in~\cite{Fan:2013yva}. As shown there, for very light $C$, $m_C \ll 1$ MeV, only a very small fraction of the total relic abundance needs to be asymmetric for cooling to happen.}. The number of $X$ ($\bar{X}$) particles captured by the Sun is governed by an equation very similar to the one discussed in~\cite{Zentner:2009is}:
\beq\label{eq:capture1}
\frac{dN_X}{dt} = C_N - C_A N_XN_{\bar{X}} + C_S (N_X+N_{\bar{X}})=C_N - C_A N_X^2 + 2C_S N_X.
\eeq
Here $C_N$ is the rate of capture by nucleons in the Sun, $C_A$ is the
rate for captured dark matter to annihilate inside the Sun, and $C_S$
is the dark matter self-capture rate computed in the the previous section. Nuclear capture rates $C_N$ have
been calculated in~\cite{Gould:1987ww, Gould:1987ir, Gould:1991hx} in terms of $\sigma_N$, the cross section for dark matter to scatter off
a nucleus. We will consider only spin-independent scattering, in which case the nucleus-DM scattering cross section is given by~\cite{Jungman:1995df}:
\beq\label{eq:form}
\sigma_{N} = \sigma_{n} A^2 \frac{(m_X m_N)^2 (m_X+ m_p)^2}{(m_X+ m_N)^2 (m_X m_p)^2}~,
\eeq
where $\sigma_n$ is the dark matter-nucleon scattering cross section, $A$ is the atomic number of the nucleus, and $m_N$ is the mass of the
nucleus. Loss of coherence is accounted for in the full formula in~\cite{Gould:1987ir, Gould:1991hx} by multiplying the cross section by an exponential nuclear form factor.  For the solar capture of heavy dark matter particles with masses above 30~GeV, the most important contribution to $C_N$ comes from dark matter scattering off oxygen atoms with $A=16$~\cite{Gould:1991hx, Zentner:2009is}. 

The annihilation rate coefficient is 
\beq
C_A = \langle \sigma_A v \rangle \frac{ \int dV n_t(r)^2}{ N_t^2},
\eeq
where $v$ is the relative velocity between annihilating $X$ and $\bar{X}$ particles.
In the dissipative dark matter scenario, dark matter particles annihilate into dark photons with a Sommerfeld-enhanced cross section
 \beq\label{eq:sigmaA}
\langle \sigma_{X \bar X \to \gamma_D \gamma_D} v \rangle &\approx& \langle \frac{\pi^2 \alpha_D^3}{m_X^2 v} \rangle \nonumber \\
&\approx&6.4 \times 10^{-23} {\rm cm}^3{\rm s}^{-1} \left( \frac{\alpha_D}{10^{-2}}\right)^3 \left(\frac{100\,{\rm GeV}}{m_X}\right)^{3/2},
\eeq
where the annihilation cross section is averaged over a Maxwell-Boltzman distribution with velocity dispersion set by $v = \sqrt{3T_\odot/m_X}$ as the captured dark matter particles are fully thermalized in the solar core before annihilation. We will validate this assumption in the appendix. 
 As shown in the previous section, captured dark matter particles only occupy a small region in the core of the Sun, in which we can approximate the density as constant, $\rho_\odot \approx 150$ g/cm$^3$~\cite{Bahcall:2004pz}. This allows us to obtain a simple analytic formula for the annihilation rate coefficient~\cite{Griest:1986yu}:
\beq \label{eq:ann}
C_A &=& \langle \sigma_{X\bar{X}\to\gamma_D\gamma_D} v \rangle \frac{V_2}{V_1^2} \approx 9.3\times 10^{-51}~\sec^{-1}\left(\frac{\alpha_D}{10^{-2}}\right)^3, \\
V_k&=&\int e^{- k M_X \phi(r)/T_\odot} d V=2.45 \times 10^{27} \left(\frac{100\,{\rm GeV}}{k\, m_X}\right)^{3/2}\, {\rm cm}^3, \quad k=1,2, \nonumber 
\eeq   
where the $V_k$'s are known as effective volumes. Note that $C_A$ only  depends on the dark coupling strength $\alpha_D$ in the formula above and not on $\sigma_n$. In deriving this formula, we assume that annihilations into SM particles are subdominant compared to annihilations into dark photons, or $\br \ll 1$ with $\br$ defined as
\beq
\br\equiv\frac{\Gamma(X\bar{X}\to {\rm SM})}{\Gamma(X\bar{X}\to \gamma_D\gamma_D)}.
\eeq
However, for large $\sigma_n \gtrsim 10^{-40}$ cm$^2$, $\br$ could be comparable or larger than 1. Then the total annihilation rate is $C_A (1+\br)$ with $C_A$ computed in Eq.~(\ref{eq:ann}). 

Notice there is an additional factor of 2 in the last term of Eq.~\eqref{eq:capture1} compared to the self-capture term in~\cite{Zentner:2009is}.  This is because the capture rates for $X$ by $X$ and by $\bar{X}$ are the same up to higher order corrections. The additional contribution to the capture of $X$ particles by $\bar{X}$ targets from $s$-channel annihilation is suppressed by $v^2 \sim T_\odot/m_X \sim 10^{-6}$ GeV/$m_X$ compared to that from Rutherford scattering. 

When the self-interactions are turned off, Eq.~\eqref{eq:capture1} has a well-known solution
\beq\label{eq:solnonint}
N_X(t) = \sqrt{\frac{C_N}{C_A}} \tanh \left(\frac{t}{\tau} \right), 
      \ \ \ {\rm with } \ \ \ \tau = (C_N C_A)^{1/2}~,
\eeq
defining the timescale $\tau$ for dark matter to reach an equilibrium
abundance in the Sun.  For ordinary cold dark matter, as long as the dark matter-nucleon
cross-section is above $\sigma_n \gtrsim 10^{-48}$ cm$^2$, this timescale
$\tau$ is much shorter than the age of the Sun, $4.7$ Gyr, and the
dark matter density in the Sun has reached a steady state with
$N_{X;{\rm eq}}=N_{\bar{X};{\rm eq}}=\sqrt{C_N/C_A}$. The total flux of SM particles, e.g., $W$ boson
pairs, resulting from dark matter annihilations is then $\Gamma = C_A N_{X;{\rm eq}}N_{\bar{X};{\rm eq}}\br=
C_N\br$.  Generically, neutrinos will be produced in the 
decay of the SM annihilation products, yielding a time-independent
neutrino flux that can be detected by neutrino telescopes such as
Super-Kamiokande~\cite{Tanaka:2011uf} and IceCube~\cite{Aartsen:2012kia}. Thus
neutrino telescopes provide an important probe of dark matter-nucleon
scattering which is complementary to direct detection experiments.

Turning on self-interactions, the solution to Eq.~\ref{eq:capture1}
becomes~\cite{Zentner:2009is}:
\beq \label{eq:steady-state2}
N_X(t)= \frac{C_N\tanh(t/\xi)}{\xi^{-1}-C_s \tanh(t/\xi)},
\eeq
where
\beq
\xi = \frac{1}{\sqrt{C_NC_A+C_S^2}}.
\eeq
When $t \gg \xi$, one obtains an equilibrium abundance
\beq
N_{X; {\rm eq}} = \frac{C_S}{C_A}+\sqrt{\frac{C_S^2}{C_A^2}+\frac{C_N}{C_A}}.
\eeq
Two interesting limiting cases are nuclear capture domination, $C_N\gg
C_S$, and self-capture domination, $C_S\gg C_N$. In the case of
nuclear capture domination, self-capture is largely irrelevant, and
the solution reduces to that in Eq.~\eqref{eq:solnonint}. In the case
of self-capture domination, the steady-state abundance becomes $N_{X;{\rm eq}}=N_{\bar{X};{\rm eq}}
\approx 2C_S/C_A$ and the flux of SM particles from dark
matter annihilation is given by
\beq \label{eq:smannrate}
\Gamma = 4\frac{C_S^2}{C_A} \br.
\eeq

As we showed in the previous section, for $\eta \lesssim 1$, the self-capture rate coefficient $C_S$ depends logarithmically on $N_X(t)$. This explicit time-dependence slightly modifies the growth of the captured dark matter particles. Yet it does not affect the the existence of a steady-state solution, and $C_S$ in Eq.~(\ref{eq:steady-state2}) should be understood as the self-capture rate when the steady state is reached. 

One can define a parameter $R_s$ that indicates which capture process dominates,
\beq\label{eq:rs}
R_s&=&\frac{C_S^2}{C_NC_A} \nonumber \\
&\approx& 0.4 \left(\frac{\rho_X}{0.2 \,{\rm GeV}/{\rm cm}^3}\right) \left(\frac{\alpha_D}{10^{-2}}\right)\left(\frac{100\,{\rm GeV}}{m_X}\right)^5\left(\frac{10^{-47}\,{\rm cm}^2}{\sigma_n}\right)\left(\frac{10^{-3}}{\bar{v}}\right)^5 \left(\frac{27}{\log\left(\sqrt{\frac{2}{3}}\frac{\bar{v}}{u_{\rm min}}\right)}\right)^2 \quad \nonumber \\
 &&\quad \quad \quad \quad \quad \quad \quad \quad \quad \quad \quad \quad \quad\quad \quad \quad \quad \quad \quad \quad \quad \quad \quad \quad \quad \quad\quad \quad \quad\quad \eta \lesssim 1 \\
&\approx& 0.6 \left(\frac{\rho_X}{0.2 \,{\rm GeV}/{\rm cm}^3}\right) \left(\frac{\alpha_D}{10^{-2}}\right)\left(\frac{100\,{\rm GeV}}{m_X}\right)^5\left(\frac{10^{-50}\,{\rm cm}^2}{\sigma_n}\right)\left(\frac{10^{-3}}{|v_{\rm rel}^\odot|}\right)^5 \quad \eta \gg 1, 
\eeq
where we approximated nuclear capture by the dominant oxygen capture. It is evident that self-capture can only dominate for relatively small nucleon scattering cross-sections.

\subsection{Solar capture of partially asymmetric dissipative dark matter} 

As discussed in Sec.~\ref{sec:review}, generic dissipative dark matter sectors will have both a symmetric relic abundance composed of equal numbers of $X$ and $\bar{X}$ ions and an asymmetric relic abundance in the form of dark atoms, i.e., bound states of $X\bar{C}$.  As the temperature inside the Sun is significantly larger than the binding energy of the dark atom, $T_\odot = 1.57\times 10^7$ K $= 1.35$ keV $\gg B_{XC} $, once dark atoms are captured and thermalized in the core of the Sun, they become fully ionized.  Thus the relevant number abundances we wish to compute are those of the dark charged particle species, $X$, $\bar X$, and $C$.  The most general $CP$-preserving equations for the solar abundance of these populations are
\beq \label{eq:generalcapture1}
\frac{d N_X}{d t} &=& C_N -  C_A N_X N_{\bar{X}}+ C_S (N_X+N_{\bar{X}}) , \\
\frac{d N_{\bar X}}{d t} &=& C_{\bar N} -  C_A N_X N_{\bar{X}}+ C_{\bar S} (N_X+N_{\bar{X}}) , \\
\frac{d N_C}{d t} &=& \tilde{C}_N + \tilde{C}_S (N_X+N_{\bar{X}}) .
\label{eq:generalcapture3}
\eeq
The rates $C_N$, $C_{\bar N}$, and $\tilde{C}_N$ are, respectively, the nuclear capture rates for $X$ ions {\em and} dark atoms, $\bar X$ ions, and dark atoms alone.  Analogously, $C_S$, $C_{\bar S}$, and $\tilde C_S$ are the self-capture rates for $X$ ions {\em and} dark atoms, $\bar X$ ions, and dark atoms alone.  

As already mentioned in Sec.~\ref{sec:selfcap}, the capture of dark atoms through self-capture (as well as nuclear interactions) is short-range in 
comparison to the dark Bohr radius $r_B$.  Thus to leading order dark atom capture occurs via the capture of the heavy particle $X$ inside the 
atom, with the light $\bar C$ following as a consequence of its dark electromagnetic interactions.\footnote{By contrast, the impact parameter for self-capture of $\bar C$ is much greater than $r_B$, and hence the dark electromagnetic interaction is screened, rendering the self-capture rate negligible in comparison.}  
The nuclear capture rates $\tilde{C}_N$ can be simply obtained from the rates for dark ions by replacing the local density of $X$ ions $\rho_{X}$ with 
that for dark atoms, $\rho_{X\bar C}$ and $C_N=\tilde{C}_N+C_{\bar{N}}$. Besides depending on a different local density, the self-capture of dark atoms also 
has a different minimal velocity that could lead to capture, $u_{\rm min}$, as demonstrated in Eq.~(\ref{eq:vcutoff}). This is mostly relevant 
for $\eta \lesssim 1$ when the relative velocity between the dark matter flux and the Sun is comparable to or smaller than the velocity dispersion.

The case when the dark atom abundance is negligible in comparison to the ion abundance, $\rho_X \approx \rho_{\bar{X}}$, is the symmetric case 
discussed in the previous section. Another interesting limit takes the dissipative dark matter to be purely asymmetric, comprised only of 
atoms, $\rho_X \ne 0,\ \rho_{\bar{X}} = 0$.  In this case there are no annihilations, and the solution to Eq.~\eqref{eq:generalcapture1}
is simply
\beq
N_X (t) = \frac{C_N}{C_S}\left(e^{C_St} -1\right),
\eeq
neglecting the possible time dependence of $C_S$, which only brings a minor modification to the numerical result. For $t\ll 1/C_S$, the amount of captured dark matter grows linearly, with the growth becoming exponential for $t \gg C_S^{-1}$.  However,  after the self-capture cross section reaches $\sigma_{\rm eff}\equiv \pi r_X^2$ at a time $t_*$, the number of captured dark atoms grows linearly again,
\beq
N_X(t)=(C_N+C_S^{\rm eff})(t-t_*)+N_X^*.
\eeq

\begin{figure}[t]
\centering
\includegraphics[width=0.6\textwidth]{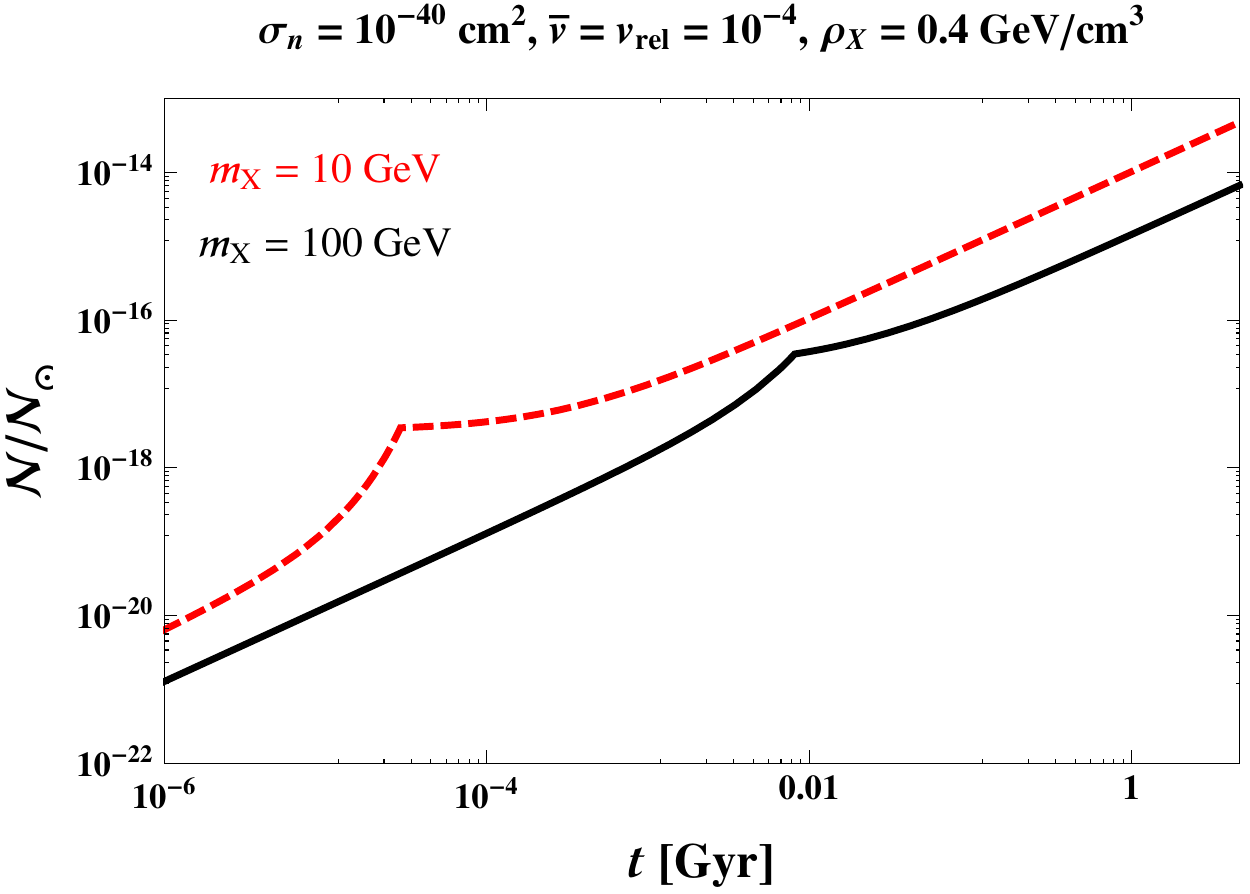}
\caption{ Number of captured dark atoms inside the Sun divided by the total number of baryons of the Sun as a function of time. We assume $\sigma_n = 10^{-40}$ cm$^2$, $\rho_X = 0.4$ GeV/cm$^3$, $\alpha_D = 10^{-2}$, $m_X = 10$ GeV, $\bar{v} = 10^{-4} c$ and $|v_{\rm rel}^\odot| = 10^{-4} c$. Red dashed curve: $m_X = 10$ GeV; black solid curve: $m_X = 100$ GeV. The kinks in both curves are the points when the geometric limit of self-capture is saturated. }
\label{fig:asymmetry_growth}
\end{figure}

Since the number density grows linearly, one might worry that the number of dark matter particles could become comparable to the number of baryons in the Sun. This is not the case.  The number of captured dark atoms as a function of time is shown in Fig.~\ref{fig:asymmetry_growth} for 10~(100)~GeV dark 
matter with $\sigma_n = 10^{-40}$ cm$^2$, $\rho_X = 0.4$ GeV/cm$^3$, $\bar{v} = 10^{-4} c$ and $|v_{\rm rel}^\odot| = 10^{-4} c$ . 
From this figure, one can see that the captured dark matter always constitutes only a small fraction of the total solar mass. Even for a big 
spin-independent DM-nucleon cross section of order of that of $Z$-exchange, $\sigma_n = 10^{-40}$ cm$^2$, the total captured dark number is about or below $10^{-13}$ of the total baryon number inside the Sun at the present day.

The general case where dissipative dark matter has both a symmetric and an asymmetric relic abundance interpolates between the two limits discussed above.  It is straightforward to solve the capture equations numerically, and our results for neutrino signals in the following section will be presented for the general case.

\section{Neutrino telescope constraints on dissipative DM in the Sun }
\label{sec:neutrino}
In Sec.~\ref{sec:direct_detection}, we have demonstrated that for dissipative dark matter with mass around or below 70 GeV, a large cross section of order of that of $Z$ exchange is still allowed by direct detection. Moreover, direct detection constraints could be totally absent if the solar system oscillates around the disk and is currently outside the dark disk. In this section we want to explore whether such large cross sections are also allowed by indirect detection experiments looking for products of dark matter annihilation inside the Sun. We will discuss constraints from neutrino telescopes, i.e, IceCube~\cite{Aartsen:2012kia}, which search for muon neutrinos from dark matter annihilation in the center of the Sun.
The flux of SM particles from dissipative dark matter annihilation is given by
\beq
\Gamma = C_A N_X N_{\bar{X}} \br.
\eeq
 In deriving all our numerical results, we use the standard solar model with parameters in~\cite{Bahcall:2004pz}. For a specific velocity distribution of dissipative dark matter, at a given mass and DM--nucleon cross section, the constraint on the flux of SM particles from dissipative dark matter annihilation is translated into a constraint on $\br (\rho_X/\rho_{\rm SHM})$ with $\rho_{\rm SHM} = 0.4$ GeV/cm$^3$. Strictly speaking, in a specific model, $\br$ is related to $\sigma_n$. For our model-independent analysis, we first treat $\br$ and $\sigma_n$ as free parameters and comment on their possible correlations at the end of this section.

\begin{figure}[!h]\begin{center}
\includegraphics[width=0.48\textwidth]{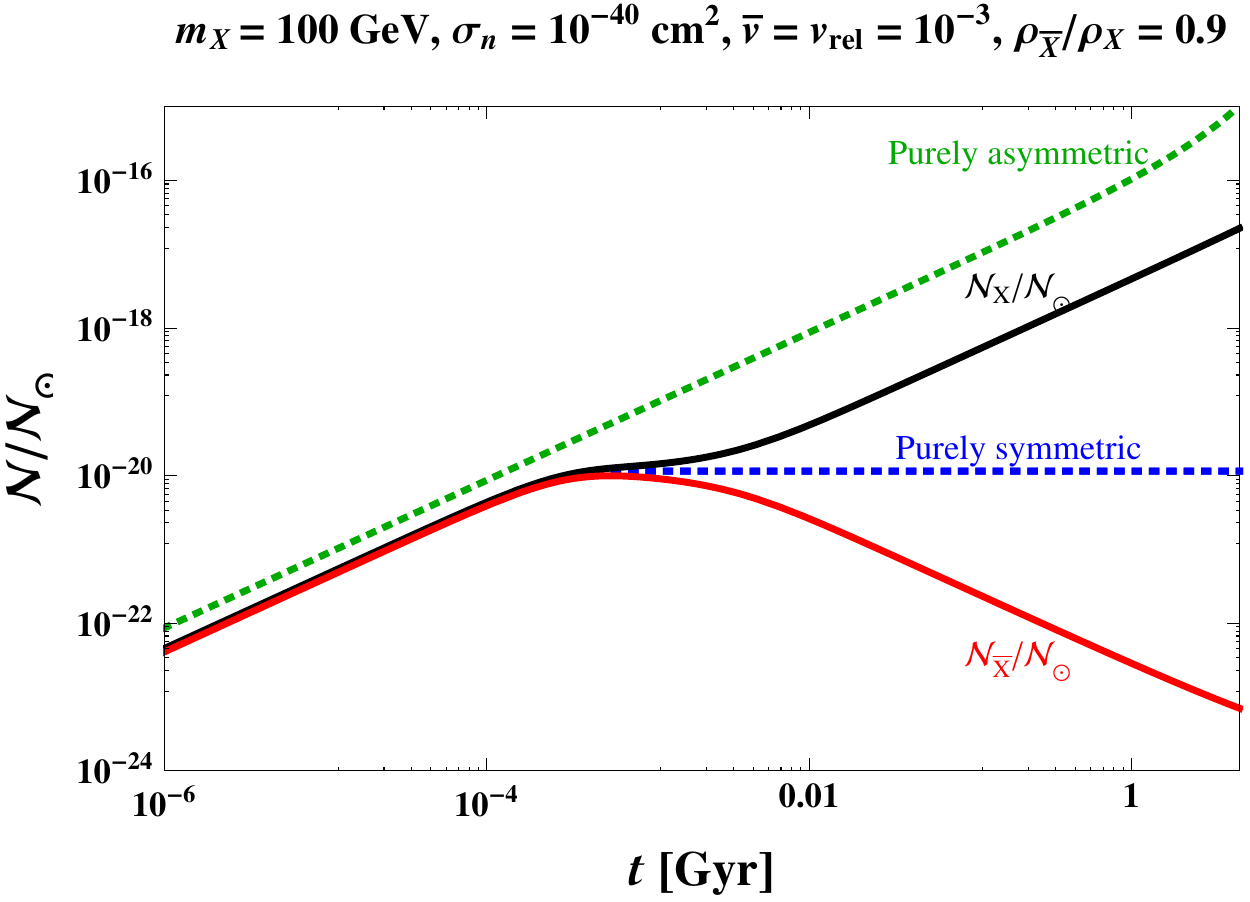} \quad \includegraphics[width=0.48\textwidth]{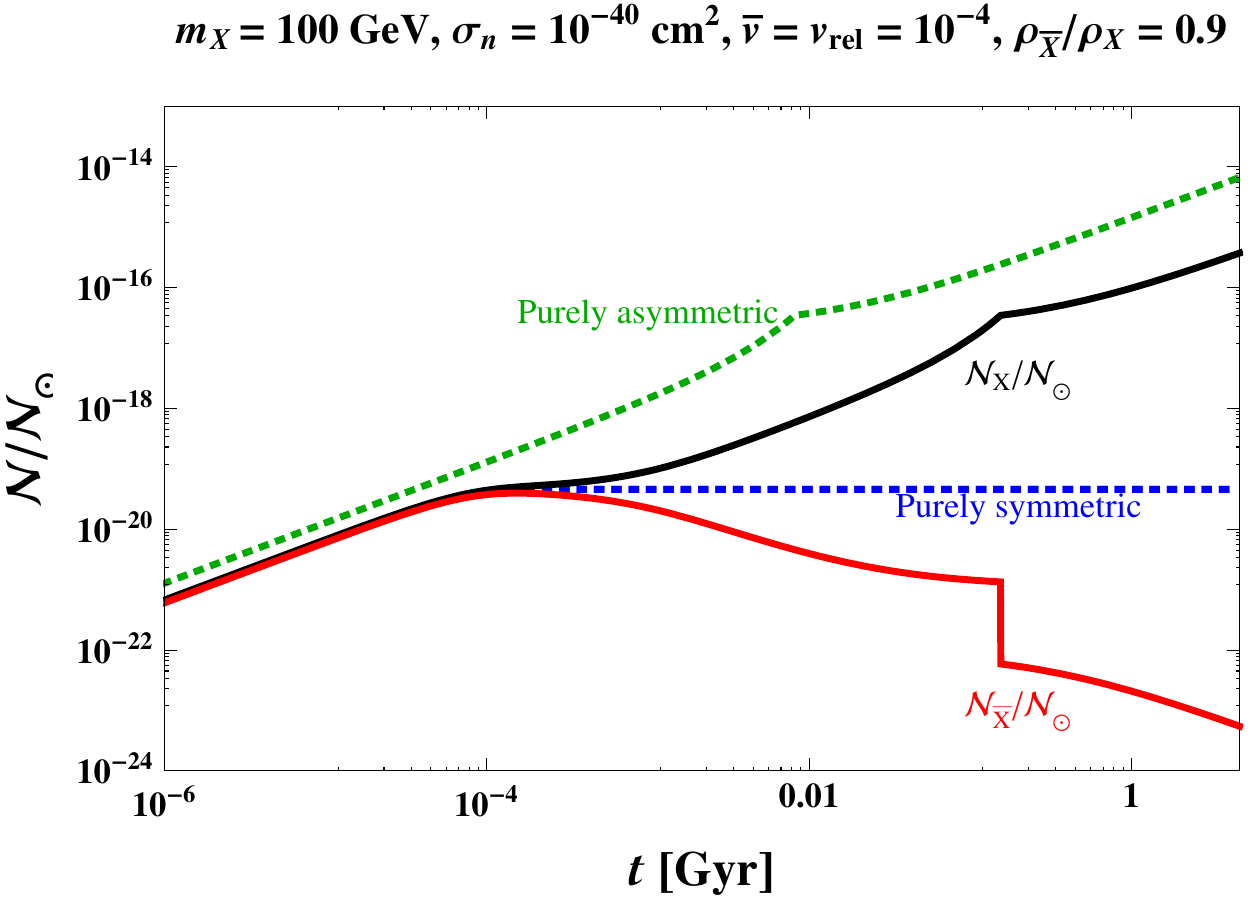} \\
\includegraphics[width=0.48\textwidth]{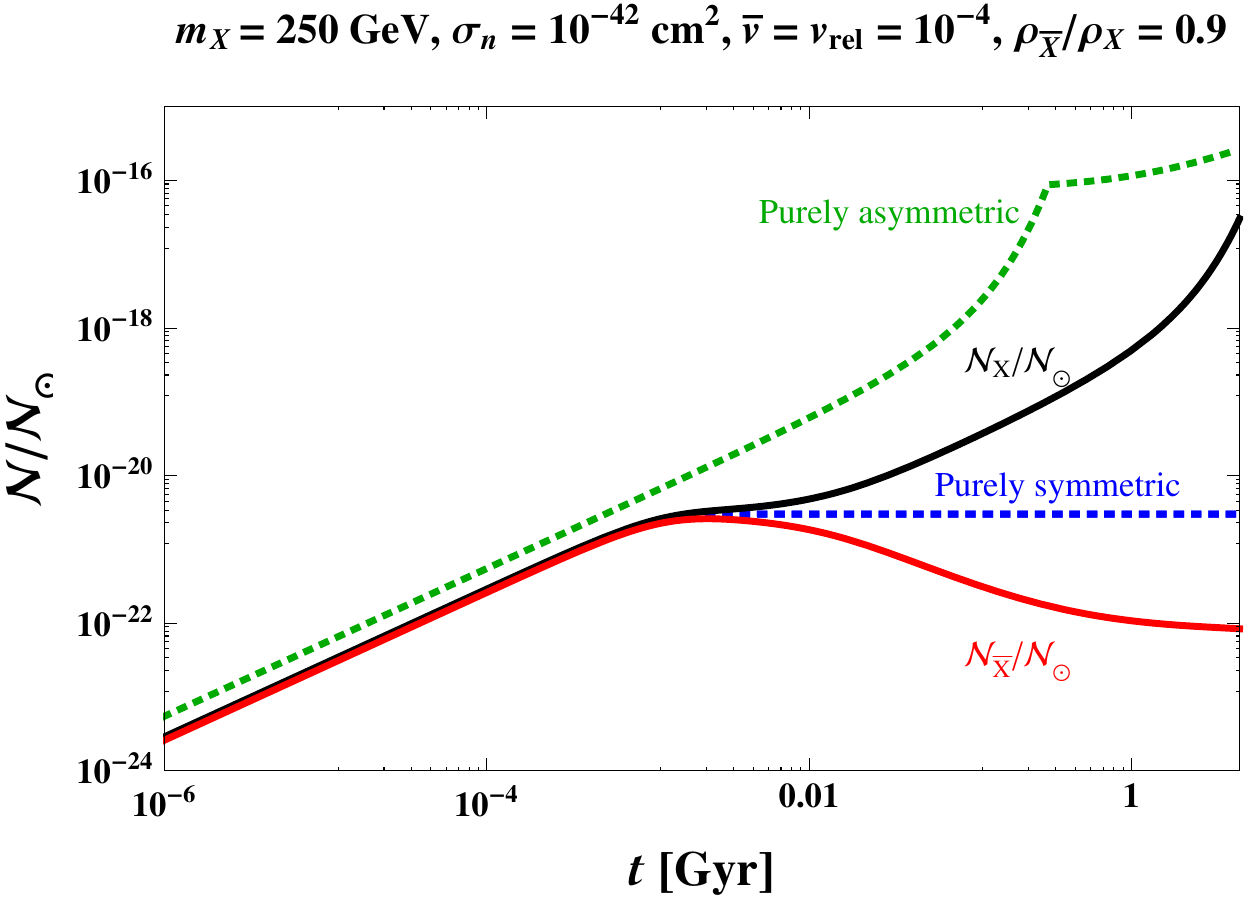}
\end{center}
\caption{Three representative histories of accumulated numbers $N_X, N_{\bar{X}}$ (normalized with respect to the number of nucleons in the Sun). For the upper two plots, we fix $m_X = 100$ GeV, $\rho_{\rm tot} = 0.4$ GeV/cm$^3$, $\rho_{\bar{X}}/ \rho_X = 0.9$, $\sigma_n = 10^{-40}$ cm$^2$. Upper left panel: $\bar{v}= v_{\rm rel}^\odot = 10^{-3}$; upper right panel: $\bar{v} = v_{\rm rel}^\odot =10^{-4}$. For the lower plot, we have  $m_X = 250$ GeV, $\rho_{\rm tot} = 0.4$ GeV/cm$^3$, $\rho_{\bar{X}}/ \rho_X = 0.9$, $\sigma_n = 10^{-42}$ cm$^2$ and $\bar{v} = v_{\rm rel}^\odot =10^{-4}$.
Black dotted curve: $N_X/N_\odot$; red dashed curve: $N_{\bar{X}}/N_\odot$. 
In each plot, we also plotted the two limiting cases corresponding to $\rho_{\bar{X}} = \rho_X$ (blue dashed curve) and $\rho_{\bar{X}} = 0$ (green dashed curve). In the upper right plots, the kinks of the black and red curves are due to the saturation of geometric limit of self-capture. In the upper right panel, the almost vertical segment of $N_{\bar{X}}$ corresponds to a brief period of exponential decrease caused by annihilation after the geometric self-capture limit is saturated.}
\label{fig:accumulationhistory}
\end{figure}

As mentioned in the previous section, a general dissipative dark matter scenario with both symmetric and asymmetric components has a solar accumulation history interpolating between the limits with a purely symmetric or purely asymmetric relic abundance. This is demonstrated in Fig.~\ref{fig:accumulationhistory}. It is evident from the figure that even with a small asymmetric DM component of order ${\cal O} (0.1)$, the accumulated dark matter particles never reach a 
steady state. $N_X$ keeps growing with the elapse of time but $N_{\bar X}$ first grows and then decreases. At the beginning, the nuclear capture 
rate is much larger than the annihilation rate, and both $N_X$ and $N_{\bar X}$ grow. Then at the time when the annihilation rate surpasses the smaller capture rate of $N_{\bar{X}}$, $N_{\bar{X}}$ starts to drop but $N_X$ keeps growing. After that, as there are always more $X$'s to annihilate with captured $\bar{X}$, $N_{\bar{X}}$ keeps being depleted. Yet the total capture rate of $X$ is always larger than the annihilation rate, so $N_X$ continues growing. The details of the growth (decrease) of $N_X (N_{\bar X})$ vary with parameters as demonstrated in Fig.~\ref{fig:accumulationhistory}. For example, in the 
upper left panel, the growth of $N_X$ is linear; in the upper right panel, the growth has a kink due to the saturation of the self-capture rate at the geometric limit, and in the lower panel, the nuclear capture rate is smaller and the growth is exponentially fast at late times. 

\begin{figure}[!h]\begin{center}
\includegraphics[width=0.48\textwidth]{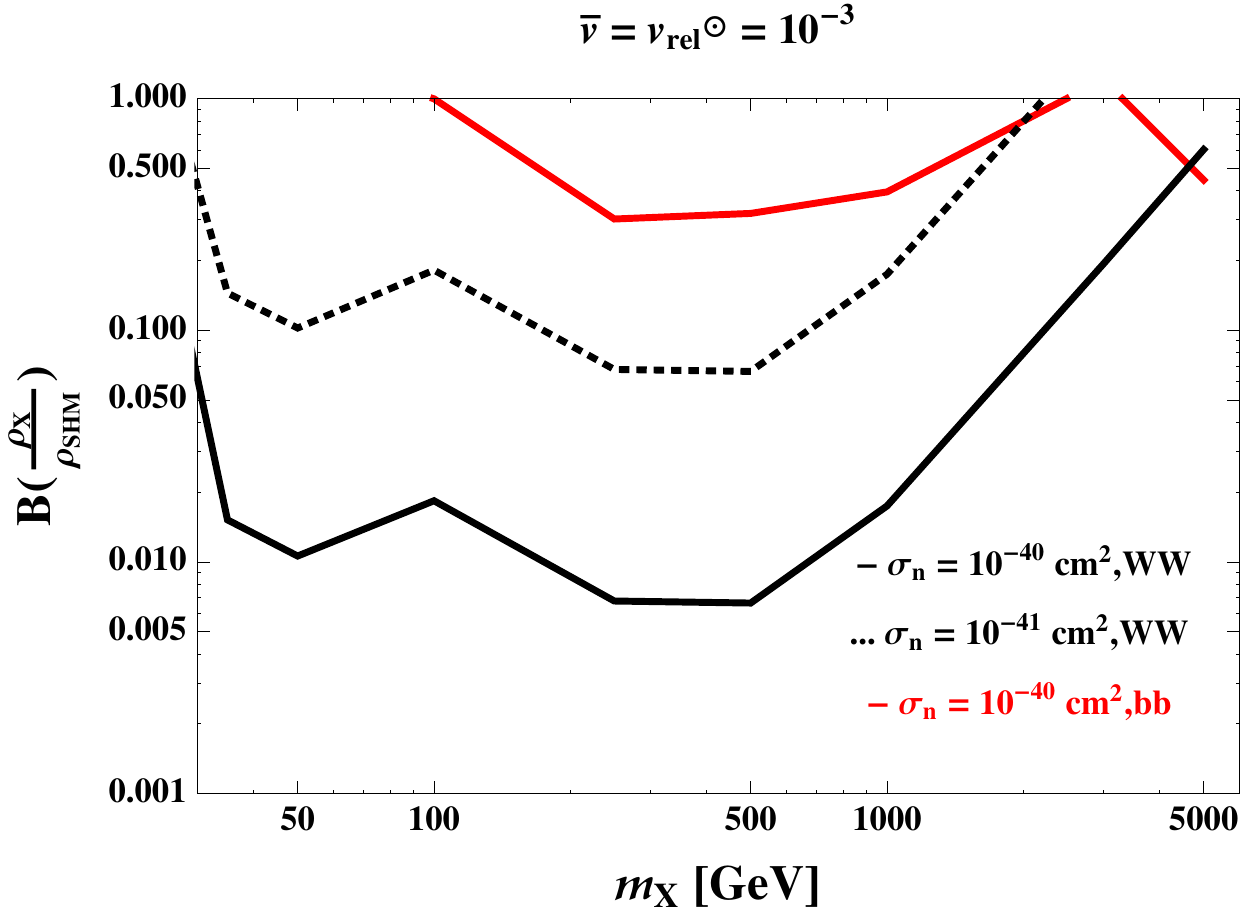} \quad \includegraphics[width=0.48\textwidth]{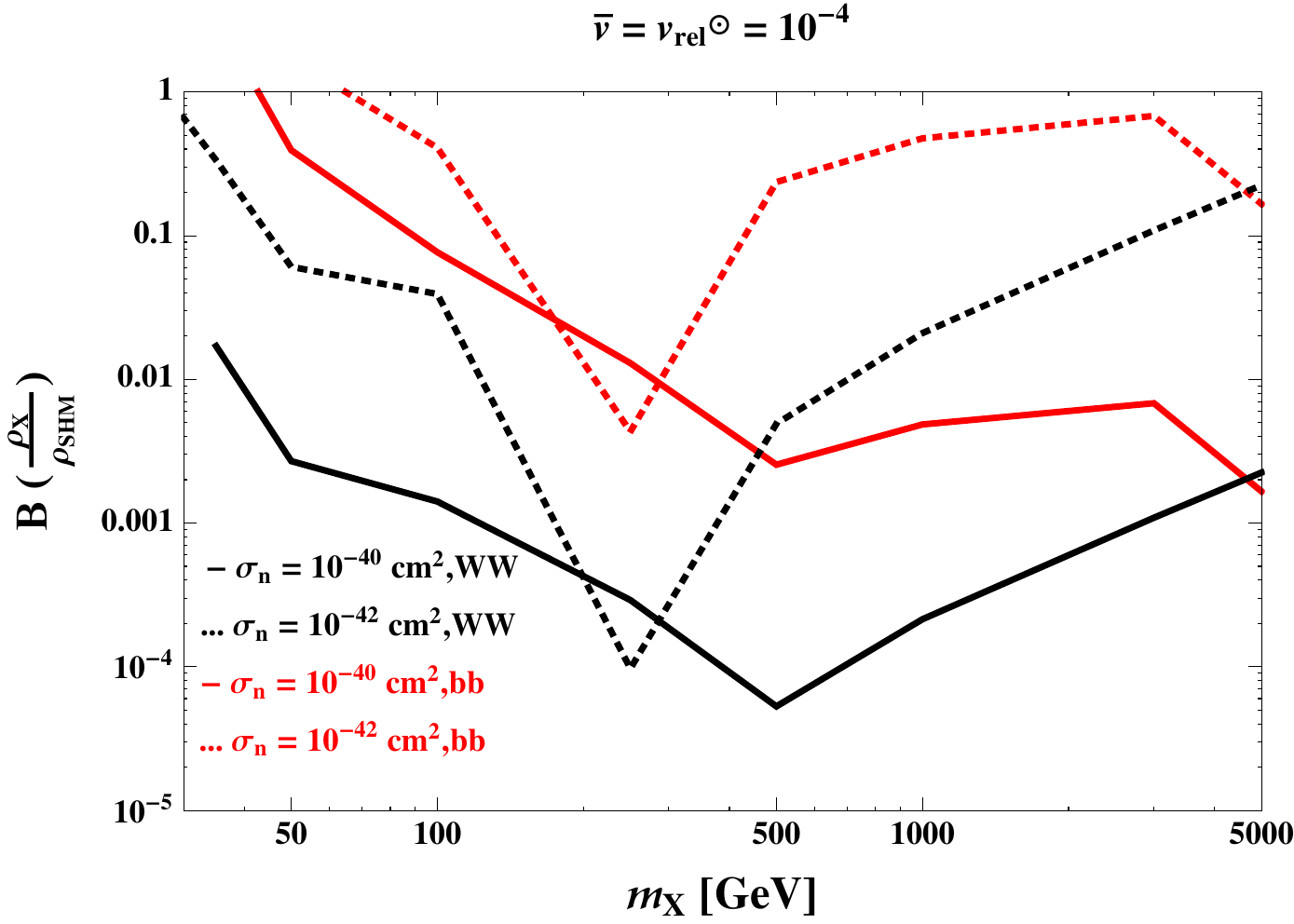} 
\end{center}
\caption{Constraints from IceCube on $\br (\rho_X/\rho_{\rm SHM})$. Left panel: $\bar{v}=v_{\rm rel}^\odot = 10^{-3}$; right panel: $\bar{v} = v_{\rm rel}^\odot =10^{-4}$. Black curves: dissipative dark matter annihilating into $W^+W^-$ ($\tau^+\tau^-$ for dark matter with mass below 80 GeV) for different $\sigma_n$'s; red curves: dissipative dark matter annihilating into $b\bar{b}$ for different $\sigma_n$'s.}
\label{fig:icecubebounds}
\end{figure}

Now we use the IceCube bounds on the solar muon neutrino flux to set bounds on dissipative dark matter annihilating into two representative classes of SM final states: $W^+W^-$ ($\tau^+\tau^-$ for lighter dark matter with mass below 80 GeV) and $b\bar{b}$. We choose two representative velocity 
parameters: $\bar{v} = v_{\rm vel}^\odot = \{10^{-3}, 10^{-4}\}$. For $\bar{v} \ll v_{\rm vel}^\odot$, all the capture rates only depend on 
$v_{\rm vel}^\odot$ and the results are unchanged.  The final results are presented in Fig.~\ref{fig:icecubebounds}. For 
$\bar{v}=v_{\rm rel}^\odot = 10^{-3}$, nuclear capture dominates as long as the cross section is not small and 
$m_X \geq 20$ GeV. For $\bar{v}=v_{\rm rel}^\odot = 10^{-4}$, the constraints get stronger due to enhanced capture rates. Self-capture could become 
important and the geometric self-capture limit will be saturated for $m_X < 500$ GeV for $\sigma_n > 10^{-42}$ cm$^2$ and for 
$m_X < 250$ GeV for $\sigma_n \lesssim 10^{-42}$ cm$^2$. 

For a large cross section $\sigma_n = 10^{-40}$ cm$^2$, we expect that $\br \sim {\cal O} (1)$ generically. To achieve such a large direct detection cross section, the dissipative dark matter could be either charged under the SM weak symmetry $SU(2)_W$ or coupled to the gluons through a dimension-seven operator $\bar{X} X G^2/\Lambda^3$ with a very low cutoff $\Lambda \sim 300$ GeV. In either case, the cross section for  annihilation into SM final states will be comparable to that of annihilation into dark photons. Thus from Fig.~\ref{fig:icecubebounds}, we see that to allow $\sigma_n = 10^{-40}$ cm$^2$, the local DDM density has to be really small, e.g., about $10^{-3}\rho_{\rm SHM}$ for mass around 100 GeV and $10^{-4}\rho_{\rm SHM}$ for a mass around 500 GeV.This is a much stronger constraint than the direct detection constraint!

Another way of interpreting Fig.~\ref{fig:icecubebounds} is that the constraint is on the symmetric fraction of the dissipative dark matter. To allow 
for $\sigma_n = 10^{-40}$ cm$^2$, the symmetric relic abundance could only be $(10^{-4} - 10^{-2})$ of the total dissipative dark matter relic 
abundance  for $m_X \gtrsim 100$ GeV if DM dominantly annihilates into $W^+W^-$ final state or $(10^{-2} - 10^{-1})$ if DM dominantly 
annihilates into $b\bar{b}$. 
 
 In summary, if there is a non-negligible symmetric component of dissipative dark matter, neutrino telescopes set much stronger constraints on the local DDDM density and/or the coupling between the dark and our visible sectors which leads to spin-independent scattering, compared to the direct detection experiments! As we mentioned in Sec~2.3 and Sec. 3, while direct detection limits could be evaded if  our solar system is currently outside the dark disk, it is much more difficult to evade the indirect detection constraints as long as our solar system has been inside the dark disk for a significant fraction of its lifetime. Thus indirect detection based on solar capture is a more robust probe of dissipative dark matter. 

\subsection{Capture in the Earth}
\label{sec:earth}

The smallness of the Earth's escape velocity, 
$\vescE = 3.74\times 10^{-5} c$, means that signals from the Earth are
proportionally sensitive to slow DM velocity dispersions to a greater degree
than solar signals.  Away from nuclear resonances, nuclear capture rates are proportional to 
$(v_{\mathrm{esc}}/\bar v)^4$  in the hard-scattering regime
relevant for capture in the Earth.  Thus for $\bar v = v_{rel}= 10^{-4} c$, the nuclear capture rate can be enhanced by as much as $10^4$ relative to
the expectation for standard dark matter.  For dark matter without appreciable
self-interactions, this enhanced nuclear capture rate translates into a correspondingly enhanced annihilation signal, potentially visible in IceCube~\cite{Bruch:2009rp}.

For dissipative dark matter, however,  self-interactions act to limit the accumulation of dark matter in the Earth.  The shallowness of the Earth's potential well means that in
collisions between an incoming DM particle and a bound DM particle,
the probability of ejecting the bound DM particle from the Earth is
significant~\cite{Zentner:2009is}.  The computation of the rate for ejection without capture is similar to that of  Eq.~\ref{eq:thetacap} describing capture without ejection.
Ejection without capture occurs when $v\equiv v^\oplus_{\mathrm{esc}}(r) < u$ and the scattering angle lies
in the range
\beq
\frac{v^2-u^2}{u^2+v^2}\leq \cos\theta \leq \frac{u^2-v^2}{u^2+v^2} \;.
\eeq
The  probability for an incoming DM particle, scattering via the Rutherford interaction, to eject the target DM particle without being captured itself is then
\beq
w\Omega _e(w) = n_t(r)\frac{\pi\alpha_D^2}{m_{DM}^2} \, \frac{u^2 - v^2}{v^2 u^2} \Theta (u^2-v^2)
\eeq
where $n_t(r)$ is the local density of target DM particles.  For the Earth, the ejection rate, $C_E$, dominates over the capture rate, $C_S$.  The evolution of DM bound to the Earth reaches a metastable equilibrium where
\beq
N_X(t) \approx N_{X0} = f_N \frac{C_N}{(C_E-C_S)}, \phantom{spacer}
 N_{\bar X}(t)\approx N_{\bar X 0} = (1-f_N) \frac{C_N}{(C_E-C_S)},
\eeq
so that nuclear capture is approximately balanced by evaporation, with the
annihilation rate $C_A N_0 \bar N_0 \ll C_E (N_0+\bar N_0), C_N$.  Here
 $f_N= \rho_N/(\rho_N+\rho_{\bar N})$ is the fraction of dissipative DM comprised of $X$ rather than $\bar X$.  This metastable equilibrium
population is much smaller than the equlibrium that would be obtained in the
absence of evaporation, and thus the net annihilation rate in the Earth is reduced to
levels which are challenging targets for current neutrino telescopes.

These conclusions were obtained for Earth interactions with the gravitationally unbound DM streaming in from outside the solar system.  This unbound population 
accounts for the majority of the dark matter
phase space density in the solar system, even for DM populations with
small velocity dispersions \cite{Bruch:2009rp}.  The Earth may also interact with the
population of DM that is gravitationally bound to the solar system
after scattering in the Sun or other planets. While an enhancement of the slow tails of the velocity distribution would increase the metastable equilibrium
population, a contribution of the degree suggested by 
numerical simulation~\cite{Peter:2009mi,
  Peter:2009mm} would not be sufficient to render the Earth signal observable.

\section{Helioseismology constraints on asymmetric dissipative DM}  
\label{sec:helioseismology}

When the dissipative dark matter is purely asymmetric, it will not be
constrained by neutrino telescopes at all. Yet the accumulation of dark
matter within the Sun could affect the solar properties studied in helioseismology
measurements.

Helioseismology measurements study the acoustic pressure waves
propagating through the Sun and yield a precise map of sound speed
over the outer 90\% of the Sun by radius~\cite{Basu:2009mi,Chaplin:2013hha}. There is a discrepancy 
between the predictions of
the standard solar model and the recently measured photospheric
metal abundances and helioseismology data, known as the ``solar
abundance problem''~\cite{Bahcall:2004yr, Basu:2004zg,
  Montalban:2004za}.  It is natural to wonder whether accumulated asymmetric
dark matter could help solve the problem. However, studies
in Ref.~\cite{Cumberbatch:2010hh} show that self-interacting asymmetric
dark matter, instead of helping to solve the problem, worsens the
discrepancy. Furthermore, Ref.~\cite{Cumberbatch:2010hh} derives a
constraint from helioseismology measurements of the core sound speed and
low-degree frequency spacing, ruling out a light self-annihilating
asymmetric dark matter candidate with mass below 10~GeV and large nuclear
cross section $\sigma_n \sim 10^{-36}$
cm$^2$.\footnote{Ref~\cite{Taoso:2010tg} reaches similar conclusions
  while Ref~\cite{Frandsen:2010yj} makes the opposite claim, that light
  self-interacting dark matter solves the solar abundance
  problem.} Given that the ``coolant'' dark matter particle must be
light in the dissipative dark matter scenario, one might worry that
helioseismology will constrain the dissipative dark matter
scenario. However, the light ``coolant'' does not have a big impact on
helioseismology as it is always bound to the heavy particle.

The basic idea why light asymmetric dark matter could affect the Sun's material properties is as follows. Assuming that dark matter with mass
$m_X$ is in thermal equilibrium with the Sun's core, it will be
concentrated in a region with scale radius $r_X$ estimated in
Eq.~\ref{eq:scaleradius}, which is about 0.1 $R_\odot \sqrt{\frac{{\rm
      GeV}}{m_X}}$.  Thus for heavy dark matter with mass $m_X > 10$
GeV, after being captured, the bound dark matter population will be localized in a small region within the core, leaving such solar properties as the radius of the
convection zone unaffected. However, for a light dark matter particle with mass in
the (sub-)GeV region, the bound population will occupy a bigger volume, and calculations
in~\cite{Cumberbatch:2010hh, Taoso:2010tg} show that the accumulated dark matter
will increase the sound speed in the inner region, thereby worsening the discrepancy between the predictions of standard
solar model and the recent solar measurements. Moreover, if the light dark matter species is lighter than
a GeV, it might evaporate unless there are other heavy dark matter
particles that attract it.

In our case, the light dark matter particle, $\bar{C}$, is always
bound into dark atoms with the asymmetric component of the heavy
particle $X$, with a binding energy around or smaller than the hydrogen
binding energy 13.6~eV in most of the parameter space that allows
cooling to happen. The dark atom is captured as a unit
and subsequently thermalizes inside the Sun's core, where the temperature is significantly larger than the binding energy of the dark atoms. If the light particles $\bar C$
expand outside the volume with size estimated in
Eq.~(\ref{eq:scaleradius}) where the heavy particle $X$ concentrates, a
charge separation would occur between the larger cloud of $C$
particles and a smaller cloud of $X$ particles. This would produce
dark electric fields that pull the $C$ particles in. Thus we expect
light $C$ particles are confined in a region similar to that occupied by the heavy $X$
particles. The separation length scale is of order the Debye
length. For instance, for 10 GeV dark matter, the final captured number is $10^{-13}
N_\odot = 10^{44}$ for $\sigma_n= 10^{-40}$ cm$^2$ as calculated in Sec.~\ref{sec:solarcap} and the
number density is $n_X = N_X/(4\pi/3 r_X^3) \approx 8 \times 10^{12}$ cm$^{-3}$. The Debye length is then of 
order $10^{-5}$ cm, much smaller than the solar radius or the scale volume radius for $X$.  Thus the overall effect 
of dissipative dark matter on helioseismology will be determined mostly by the heavy particle $X$. As long 
as the cross section of $X$ scattering off nucleons is smaller than $10^{-36}$ cm$^2$ and/or $m_X > 10$ GeV, the helioseismology 
constraints discussed in~\cite{Cumberbatch:2010hh, Taoso:2010tg} do not apply to our case.

\section{Conclusions and Outlook} 
\label{sec:conclusion}

In this article, we consider the experimental constraints from direct
and indirect detection experiments on dark matter scenarios with a
dissipative component.  The dissipative dark matter could cool through
mechanisms similar to baryons and form collapsed dark structures such
as a dark disk. The relic abundance will generally consist of both a
symmetric and an asymmetric component.  We systematically work out the
theory of solar capture for dissipative dark matter, and evaluate the
resulting constraints from dark matter annihilation in the Sun as well
as from direct detection.

We first demonstrate that due to the novel spatial and velocity
distributions of dissipative dark matter, the limits on its couplings
to SM particles and its local density can be evaded or relaxed
considerably compared to that of normal cold dark matter. However, if
the symmetric component of the relic abundance is non-negligible,
neutrino telescopes looking for annihilation products of dark matter
captured by the Sun set much stronger constraints. For dark matter
cooled to the minimum interesting velocity dispersions of order
$10^{-4}$, a dissipative dark matter mass of 500~GeV and a cross
section for scattering off nucleons as large as $10^{-40}$ cm$^2$,
either the fractional symmetric abundance or the (integrated) local
density of dissipative dark matter in the solar system compared to
that of normal cold dark matter has to be less than $4 \times
10^{-5}$.

For light purely asymmetric dissipative dark matter, with mass in the
GeV range, accumulation in the solar core could affect
helioseismological data. So far helioseismology only limits a
DM--nucleon cross section of order $10^{-36}$ cm$^2$.  IceCube has
very little sensitivity to dark matter with masses $\lesssim$ 15 GeV,
but lower-energy neutrinos from annihilation products showering inside
the Sun would allow a lower-threshhold experiment such as
Hyper-Kamiokande to observe a signal in this regime
\cite{Rott:2012qb}.  We also consider Earth capture and show that,
unlike the case of solar capture, DM self-interactions act to limit
the accumulation of dark matter in the Earth and thus reduce the
discovery potential.

Here we have focused on the case where dissipative dark matter is
fermionic. In both fermionic and bosonic scenarios, the accumulation
of DM particles inside neutron stars might lead to black hole
formation, which could potentially set interesting bounds. This effect
has been studied in the context of normal cold asymmetric dark matter
in~\cite{deLavallaz:2010wp, Kouvaris:2010jy, McDermott:2011jp,
  Kouvaris:2011fi, Kouvaris:2011gb, Kouvaris:2012dz, Bramante:2013hn,
  Bell:2013xk, Bertoni:2013bsa, Bramante:2013nma}.  We leave the study
of this effect in dissipative PIDM models for future work.

\acknowledgments

We thank Lisa Randall for collaboration in the early stages of this
project. We are grateful to Matt Reece for numerous useful discussions
and Fabio Iocco and Eric Kramer for helpful comments. We are supported
in part by the Fundamental Laws Initiative of the Harvard Center for
the Fundamental Laws of Nature.


\appendix
\section{Consistency checks}
One key assumption that we have relied on extensively in our analysis
of the solar capture is that after being captured, the dark matter
particles quickly thermalize with the the solar core. In this
appendix, we will demonstrate this is indeed the case by showing that
the thermalization rate is faster than the rate of energy inflow due
to self-capture as well as the the annihilation rate provided that the
the cross section of scattering between dark matter particles and
nucleons is not tiny.

The energy transfer rate per volume during thermalization is approximately
\beq
\frac{d E_{\rm thermal}}{dt dV} &\approx & n_p n_t \sigma_n v_{\rm esc}^\odot \langle \Delta_E \rangle \nonumber \\
&\approx & n_p n_t \sigma_n \frac{m_Xm_p}{(m_X+m_p)^2} m_X v_{\rm esc}^{\odot 3},
\eeq
where $n_p$ is the number density of nucleons at the solar core, $n_p \approx 8 \times 10^{25}$ cm$^{-3}$. We neglected the velocity distribution and approximated the kinetic energy of dark matter particles after being captured to be $E_{\rm kin} \sim m_X v_{\rm esc}^{\odot 2}$ and the relative velocity between dark matter particles and nucleons to be $v_{\rm esc}^\odot$. The energy transfer between dark matter particles and nucleons is $(m_Xm_p/(m_X+m_p)^2)$ fraction of $E_{\rm kin}$. Similarly the rates of energy inflow due to self-capture and annihilation are respectively
\beq
\frac{d E_{\rm self}}{dt dV} &\approx & n_X n_t \langle\sigma_{\rm cap} \rangle m_X v_{\rm esc}^{\odot 3}, \\
\frac{d E_{\rm ann}}{dt dV} &\approx & n_X n_t \langle\sigma_{A} \rangle m_X v_{\rm esc}^{\odot 3}.
\eeq

The requirement that the thermalization rate is faster than the self-capture  and annihilation rates then  translates to 
\beq
\frac{\sigma_n}{\langle\sigma_{\rm cap} \rangle}, \frac{\sigma_n}{\langle\sigma_{A} \rangle} > \frac{\rho_X}{m_p n_p} \approx 5 \times 10^{-27}\frac{\rho_X}{0.4 \, {\rm GeV}/{\rm cm}^3}.
\eeq
Given Eqs.~(\ref{eq:avexseccap}) and (\ref{eq:sigmaA}), this requirement can be easily satisfied. For instance, for $m_X = 100$ GeV and $\alpha_D =10^{-2}$, thermalization is the quickest process as long as $\sigma_n > 10^{-50}$ cm$^2$. 

Another hidden assumption in our solar capture analysis is that the dark photons produced from dark matter annihilation do not reheat the dark plasma in the solar core. The easiest way to check this is to compute the mean free path of the dark photon, which is given by 
\beq
\ell &=& \frac{1}{n_t \sigma_{\rm Comp}} = \frac{3 m_X^2}{n_t 8\pi \alpha_D^2} = \frac{m_X^2 r_X^3}{2 N_t \alpha_D^2} \nonumber \\
&\approx & 10^{13}\,{\rm km} \sqrt{\frac{m_X}{100 \, {\rm GeV}}} \frac{10^{-13} N_\odot}{N_X} \left(\frac{10^{-2}}{\alpha_D}\right)^2,
\eeq
where in the first line $\sigma_{\rm Comp}$ is the Compton scattering cross section between captured dark matter and the dark photons, $\sigma_{\rm Comp} =  (8\pi\alpha_D^2)/(3m_X^2)$. It is clear that the mean free path of dark photons is much larger than the solar radius and thus they travel all the way outside the Sun freely without disrupting the captured dark plasma inside the Sun.

\bibliography{lit}
\bibliographystyle{jhep}
\end{document}